\documentclass{article}[11pt]

\usepackage[english]{babel}

\usepackage[a4paper,top=2cm,bottom=2cm,left=3cm,right=3cm,marginparwidth=1.75cm]{geometry}

\usepackage{amsmath,amssymb,amsthm}
\numberwithin{equation}{section}
\usepackage{graphicx}
\usepackage[colorlinks=true, allcolors=blue]{hyperref}
\usepackage{braket}
\usepackage{bbold}
\usepackage[dvipsnames]{xcolor}
\usepackage{tikz-cd}
\usepackage{caption}
\usepackage{cancel}
\usepackage{comment}
\usepackage{esint}
\usepackage{cleveref}
\usepackage{wasysym}
\usepackage[normalem]{ulem}
\usepackage{pifont}
\usepackage{bbm}

\begin{document}
\begin{titlepage}
\begin{flushright}
LMU-ASC 12/25\\
\vspace{1cm}
{\it{Something's got to give}}\\
{\small English proverb}
\par\end{flushright}
\vskip 1.5cm
\begin{center}
\textbf{\Large \bf BCOV on the Large Hilbert Space}
\vskip 1cm
\vskip 0.5cm
\large {
\bf 
E.~Boffo$^{a,}$\footnote{boffo@karlin.mff.cuni.cz}
, O. Hulik$^{b,}$\footnote{ondra.hulik@gmail.com} 
and I. Sachs$^{c,}$\footnote{Ivo.Sachs@physik.uni-muenchen.de}
}

\vskip 1cm

~
\\
{$^{(a)}$ \it Faculty of Mathematics and Physics, Mathematical Institute, \\ 
 Charles University Prague, Sokolovsk\'{a} 83, 186 75 Prague}
\\

{$^{(b)}$} \it Institute for Mathematics 
 Ruprecht-Karls-Universitat Heidelberg,
\\ 69120 Heidelberg, Germany
\\

{$^{(c)}$ \it Arnold-Sommerfeld-Center for Theoretical Physics, Ludwig-Maximilians-Universit\"at \\ Theresienstr. 37, D-80333 Munich, Germany}

\end{center}

\vskip 1.5cm

\begin{abstract}
We formulate the BCOV theory of deformations of complex structures as a pullback to the super moduli space of the worldline of a spinning particle. In this approach the appearance of a non-local kinetic term in the target space action has the same origin as the mismatch of pictures in the Ramond sector of super string field theory and is resolved by the same type of auxiliary fields in shifted pictures. The BV-extension is manifest in this description. A compensator for the holomorphic 3-form can be included by resorting to a description in the large Hilbert space.  

\end{abstract}

\end{titlepage}

\tableofcontents

\section{Introduction}
The idea of constructing a field theory that could give rise to string amplitudes dates back to Siegel \cite{siegel1985covariantly}. For various string theories it is known how to associate a perturbative BV-action in the form of an $L_\infty$ structure for off-shell strings \cite{WITTEN1986291}, \cite{Zwiebach:1992ie}. Likewise for the string B-model, whose target space field content are the deformations of complex structures, there is a well established history:  \cite{Bershadsky:1993cx,Barannikov:1997dwc,Costello:2019jsy}, to cite just a few of the many important contributions, as well as the recent \cite{Raghavendran:2024xqn} on the minimal model and \cite{Ben-Shahar:2025dci} for a study in the context of the double copy. The field theory resulting from the seminal work of Bershadsky--Cecotti--Ooguri--Vafa \cite{Bershadsky:1993cx} is a celebrated result and it has been named BCOV after the authors. 

{Moreover the complex modulus is related by mirror symmetry \cite{Candelas:1990rm} to the K\"{a}hler modulus, which also admits a field theory capturing its deformations \cite{Bershadsky:1994sr}}. Coincidentally, a powerful framework to deal with mirror symmetry happens to be what is commonly referred to as topological string theory, that is a sigma model on maps$(\Sigma, M)$ with a global $\mathcal{N}=(2,2)$ supersymmetry on the source space \cite{Gates:1984nk}, twisted to decouple local super reparametrizations of the worldsheet \cite{Witten:1988xj}. Depending on the twist of the supercharges (A or B), this sigma model is then sensitive either to the complex structure, or to the K\"ahler modulus for the target space. In particular, in the B-twist, (part of) the BRST cohomology 
is isomorphic to the space of equivalence classes of linear deformation of complex structures of a Calabi--Yau manifold. Linear deformations are parametrized by the Beltrami differentials as antiholomorphic 1-forms, with values in the holomorphic tangent bundle. While the original Kodaira--Spencer theory of deformations of complex structures (e.g. \cite{Schnell}) is well defined on any complex manifold, the more restrictive CY-structure is required to formulate a topological string theory with such a target space.

All of this suggests that the Kodaira--Spencer theory of complex structure deformations could be a natural playground to explore string field theory in this setting and this was indeed discussed in the original BCOV description of Kodaira--Spencer gravity  \cite{Bershadsky:1993cx}. On the other hand, one could reverse the question and ask if string field theory could help us sharpen our understanding of BCOV and, in particular, the complication due to the presence of a non-local kinetic term. 
The latter is implied by the absence of shifted symplectic pairing for holomorphic polyvector fields that allows to construct the canonical Maurer--Cartan action functional.

In this note we point out a close relation with the problem of \emph{pictures} in the Ramond sector of super string field theory. Pictures label the degree superforms, i.e.~differential forms on the moduli space of the super string world sheet. Concretely, we will import a mechanism  from string field theory \cite{Konopka:2016grr} by adding auxiliary fields with shifted picture.

In fact, as we will point out, a worldline (rather than a worldsheet) sigma-model is sufficient to construct such an action.  
Indeed, since in the B-model Maps$(\Sigma, M)$ localize on constant maps in the target space $M$, one should also be able to capture the field theory from the first quantized worldline. This leads to some technical simplifications for the super ghost sector which has a less intuitive  representation in worldsheet conformal field theory. Below, in \cref{sec:models}, we will construct a BV-action for KS-gravity theory starting from the spinning worldline with\footnote{We use the semicolon in order to emphasize that these are not left- and right moving SUSY's.} $\mathcal{N}=(2;2)$ supersymmetry on the source space, where, in contrast to the standard construction, one of the supersymmetries is gauged. This brings up a \emph{picture changing operator} that is equivalent to the divergence operator on polyvector fields whose inverse enters in the kinetic term of BCOV. Then, introducing potentials of picture shifted by $-1$, {this results in an action for complex structure deformation with auxiliary fields and a local kinetic term}. Upon symplectic reduction, we reproduce the BV-extension of the Kontsevich--Barannikov action \cite{Barannikov:1997dwc} with an odd degenerate $(-1)$-shifted Poisson bracket. The construction is fairly closely related to that in the Ramond sector of string field theory although the gauge sector looks different. An additional feature, common in string field theory, is that the existence of such an action functional implies a natural  pairing of odd degree together with the complete BV-spectrum of fields and anti-fields.   

There exists an extension of BCOV with the inclusion of an extra function, $g$ that plays the role of a compensator to ensure the holomorphicity of the Calabi--Yau 3-form \cite{Costello:2019jsy}. This function is not accounted for in the discussion just described. We will then consider an alternative formulation in \cref{sec:lhs}, which is reminiscent of the \emph{large  Hilbert space} in string theory 
\cite{Dixon:1986qv,Belopolsky:1997bg}, where the superghost sector is represented by a Laurent series with a novel, even pairing on the super ghost sector. There, the ghost module is very much suggestive of the gravitational descendant of cyclic cohomology \cite{Barannikov:1997dwc}\cite{Costello:2012cy}. Let us briefly recall how the story goes there: the field content of the B-model string field theory consists of holomorphic polyvectors. In fact, the starting point is Hochschild cohomology but the identification with polyvectors is the isomorphism in Hochschild--Konstant--Rosenberg theorem. Topological string field theory is instead related to cyclic Hochschild cohomology and therefore derived invariants of the circle action. Cyclic cohomology is a Lagrangian in periodic cyclic cohomology. By work of Melani--Safronov \cite{Melani_2018,Melani_2018_2}, any Lagrangian of a $(-n)$-shifted symplectic space has a $(n-1)$-shifted Poisson structure. Butson--Yoo \cite{Butson:2016jys} showed that the converse is also true. The cyclic cohomology of the topological SFT has an even symplectic form induced by the Calabi--Yau structure, and in turn BCOV theory has an odd Poisson bracket.  A similar story holds true in \cref{sec:lhs}, but there will also be some differences. So in the large Hilbert space we can include the compensator $g$ in the multiplet at the price of redefining the structure of the multiplets. The final result is the BV-formulation  of \cite{Costello:2019jsy} up to cubic interactions and complemented with a local kinetic term. However, as in \cite{Bershadsky:1993cx,Costello:2019jsy} the BV-bracket involves a constrained variation of the antifields. 

Let us compare our findings with literature: As will become clear in the the body of this paper, both, in the small and the large Hilbert space, our formulation including auxiliary multiplets is closely related to the \emph{theory of potentials}. As such our symplectic pairing can be made non-degenerate but the induced BV-bracket, to represent the gauge invariance of our action, is non-invertible. So, effectively we are describing a degenerate BV theory in accordance with \cite{Costello:2019jsy} for example.
 
Another feature of the worldline, in contrast to the string, is that it can be quantized on any complex manifold. So one may wonder if we can formulate a target space field theory on a generic Kähler manifold that is not necessarily Calabi--Yau. In \cref{sec:backgr-fields} we will argue that this can indeed be done, however in a different formulation which is often referred to as a ``theory of background fields" (e.g. \cite{Grigoriev:2021bes}). Here, the integrability conditions for complex structure deformations are implied by the nilpotency of the BRST differential. This is then a background independent formulation of BCOV. However, the operator state correspondence between deformation of the BRST differential and perturbative states in BCOV still requires a holomorphic 3-form defined at least locally.

A natural objection to our formulation is that the ``extended deformation space" \cite{Barannikov:1997dwc,Gualtieri:2003dx} requires a $2$-form on the source space\footnote{As communicated to us by M.Zabzine.} via the AKSZ-construction  \cite{Pestun:2006rj} which cannot be pulled back to a worldline. However, it can be pulled back to the $\mathcal{N}=2$ spinning world line so that this is not an obstruction. Another manifestation of this is that the equations of motion for a Kalb-Ramond field can indeed be derived from the world-line.  

Two appendices are attached to this manuscript. In the first one, some fundamentals on the deformation theory of complex structures are recalled, while \cref{BRST-22} discusses, for the curious reader, a generalization of the worldline model of \cref{local20}.

\section{Set-up}

Kodaira--Spencer or BCOV is the field theory on the target space of closed topological strings. 
The worldsheet theory is known as B-model, see \cite{Witten:1988xj} and \cite{Alexandrov:1995kv}. However, since $\text{Maps}(\Sigma, M)$ localize on points in the target manifold \cite{Witten:1988ze}, it is feasible to directly construct a particle model with superdiffeomorphism invariance (though strictly speaking invariance under diffeomorphisms is trivialised/removed) on the $\mathcal{N}=(2;2)$ superline and still obtain the very same field theory \cite{Marcus:1994em}. This is the perspective that we adopt in this article, and we illustrate several variants of it in section \ref{sec:models}. For the moment being, we would like to give a non-exhaustive review on Kodaira--Spencer theory (while Appendix \ref{App} refreshes the reader on the topic of complex structure deformations and at the same time collects useful formulas and identities for multivector fields).

\subsection{Review of BCOV formulations}\label{sec:rev}

We begin with a brief review of BCOV theory \cite{Bershadsky:1993cx} and its extension in  \cite{Barannikov:1997dwc} and \cite{Costello:2012cy}, \cite{Costello:2019jsy} in a fixed number of dimensions. The objective is an action functional {over a Calabi--Yau three-fold $CY^3$.} {Its} linearized field equations modulo gauge symmetries capture the cohomology
\[
H^\bullet_{\bar\partial}(\ker \text{div} \subset PV^{\bullet,\bullet}),  
\]
where the differential operators $\bar\partial$ and $\text{div}$, act on $PV^{\bullet,\bullet} := \Omega^{(0,\bullet)}(M, \Lambda^\bullet TM_{(1,0)})$ 
(see Appendix \ref{App}). The interactions arise from a 2-product given by the Schouten--Nijenhuis bracket  
\[
[-,-]_{SN} : PV^{i,j} \wedge PV^{k,m} \to PV^{i+k-1,j+m}.
\]
Tian \cite{Tian1987SmoothnessOT}, in joint work with Bogomolov and Todorov, showed that the Schouten-Nijenhuis bracket is a derived bracket with the divergence operator \eqref{TIAN}. In particular, it can boil down to
\[
[\alpha,\beta]_{SN} = \text{div} (\alpha \wedge \beta), \quad \alpha, \beta \in \ker \text{div} \subset PV^{\bullet,\bullet}.
\]
In addition, BCOV requires a trace on polyvectors defined by
\[
\int_{CY^3} \alpha : = \int_{CY^3} \Upsilon \wedge \iota_\alpha \Upsilon, \quad \alpha \in PV^{3,3} , \quad \Upsilon \in \Omega^{3,0}_{\text{hol}}(CY^3),
\]
with the properties (if the CY manifold has no boundary): $\int \alpha \bar\partial \beta = -(-1)^{\vert \alpha \vert} \int \bar\partial \alpha \wedge \beta$, and $\int \alpha \wedge \text{div} \beta = -(-1)^{\vert \alpha \vert}\int \text{div} \alpha \wedge \beta$. The latter is due to the linearity and graded symmetry of the inner derivation $\iota_{\alpha \wedge \beta} = \iota_\alpha \iota_{\beta} = \pm \iota_\beta\iota_\alpha$, and the defining property that relates div to $\partial$ (see \eqref{comm_diagram} in \cref{App}).

Consider now $\mu \in \ker \text{div} \subset PV^{1,1}  $ a Beltrami differential and $\hat{A} \in \mathcal{H}$ its harmonic component. Kodaira--Spencer theory, as described in the seminal work  \cite{Bershadsky:1993cx}, is then described by the action functional
\begin{equation}
    S_{BCOV}[\mu] = \frac{1}{2} \int_{CY^3}  \mu \wedge \frac{1}{\text{div}} \bar\partial \mu +\frac{1}{3} (\mu + \hat{A})^{\wedge^3} .
\end{equation}
Because of the $\partial\bar \partial$-lemma, there exists a $v\in  PV^{2,1}$, such that $\iota_{\bar\partial \mu}\Upsilon  = \partial \iota_{ \bar\partial v}\Upsilon = \iota_{\text{div}\bar\partial v} \Upsilon $. So we can make the following choice
\begin{equation}
\frac{1}{\text{div}} \bar\partial \mu  = \bar\partial v, 
\end{equation} 
which implies that the action is well-defined despite its manifestly non-local kinetic term. Another feature to stress is that BCOV requires a holomorphic volume form so that K\"{a}hler geometry is not sufficient (see, however,  \cref{sec:backgr-fields} for considerations on general K\"{a}hler geometries).
$S_{BCOV}$ has gauge symmetries given by
\[
\delta \mu = \bar\partial \epsilon + [\epsilon, \mu]_{SN}, \quad \epsilon \in \ker \text{div} \subset PV^{1,0},
\]
which can be verified using the derivation property of $\bar\partial$ and $\text{div}$ and Jacobi identity for $[-,-]_{SN}$. In \cite{Bershadsky:1993cx}, the BV formulation of the theory was also presented. We comment only on the classical part of it ---the antifields and the odd BV bracket. One way to achieve a BV formulation on a CY 3-fold is by determining that
\begin{equation}
\begin{cases}
   \oplus_{i+k\leq 2} PV^{i,k}  & \text{are fields},\\
   \oplus_{m+n>2} PV^{m,n} \; \cap m\neq3 & \text{are antifields}.
\end{cases}
\label{fld-antifld}
\end{equation}
Note that this assignment returns precisely an even dimensional manifold (cotangent manifold to the space of fields). Then the BV bracket, 
effectively picking the arguments $F,L \in PV^{3,3}$ of functionals denoted by the same letter, is given by
\begin{equation}
  {{\left\{ F, L\right\}}} =  \sum_{\Phi} \int_{CY^3} \Upsilon \wedge \Big(\langle   \langle \text{div}  \frac{\delta_R F}{\delta \Phi},  \Upsilon\rangle , \frac{\delta_L L}{\delta \Phi^*}\rangle - \langle  \langle \text{div}  \frac{\delta_R F}{\delta \Phi^*}  ,  \Upsilon\rangle , \frac{\delta_L L}{\delta \Phi}\rangle\Big).
  \label{BV-br}
\end{equation}
Here, pointed brackets refer to the canonical contraction of polyvectors with complex forms. The bracket is degenerate therefore one cannot associate it with a symplectic form.

Let us now take the Beltrami differential $\mu = \text{div} A \in \text{im} \, \text{div} \subset PV^{1,1} $. Then, Kodaira--Spencer theory as found in Barannikov--Kontsevich \cite{Barannikov:1997dwc} is given by
\begin{equation}
    S_{BCOV}[A]= \frac{1}{2} \int \bar\partial A \wedge \text{div} A + \frac 1 3 (\text{div}A + \hat{A})^{\wedge^3}.
    \label{BK}
\end{equation}
Both of the functionals for the classical fields \eqref{BK} and the BV functional (obtained by considering the set of polyvectors in \eqref{fld-antifld}) capture only deformations of complex structures. Instead, there is no room for a scalar compensator required to ensure that the new volume form is again holomorphic, see \cref{g-compensates-vol}.

An observation due to Costello and Li \cite{Costello:2012cy,Costello:2019jsy} fills this gap in an elegant, unified framework which was crucial in order to extend the B-model to higher genera. They work in a $CY$ of arbitrary dimension $n$ and consider instead the module 
\begin{align}
 \mathbb{C}[[u]] \otimes PV^{\bullet,\bullet}[2]\,,
\end{align}
where $u$ carries ghost degree 2, so that the total ghost number of a generic element $u^n \omega^{(i,j)}$ is
\begin{align}\label{eq:ghC}
\text{gh}(u^n \omega^{(i,j)}) = 2 - ( i + j + 2 n)\,.
\end{align}

Furthermore, it is possible to combine the differentials into an equivariant operator of ghost degree $-1$
\[
\bar\partial + u \text{div} =: Q.
\]
Thus, at order $u^0$, {the Schouten bracket} becomes the derived bracket of $Q$ with exterior multiplication of polyvectors. Then the following Maurer--Cartan equation encompasses both the equation for the deformation of complex structures and that for the preservation of the holomorphic volume form 
\begin{equation} Q a + \frac 1 2 [a,a]=0, \quad \text{gh}(a) =0.\label{MC-CostelloLi}\end{equation}
Indeed, when $a$ has ghost degree zero and $a = \mu + u g$, the equation splits into 
\begin{align}\label{eq:ELCostello}
    \bar \partial\mu+\frac{1}{2}[\mu,\mu]&=0,\\
    u(\bar \partial g+\mathrm{div} \mu+[\mu,g])&=0 . \label{eq:ELCostello2}
\end{align}
The form \eqref{MC-CostelloLi} is particularly nice because the gauge symmetries can be found right away in ghost degree $+1$. However it was not possible to come up with an action functional that has the above MC equations as e.o.m.'s. Instead Costello and Li suggested an interaction term $S_{int}$ that satisfies the BV master equation with $Q$ and the bracket given in \eqref{BV-br}
\[
Q S_{int} + \frac 1 2 \{S_{int},S_{int}\} = 0.
\]
The expression for $S_{int}$ is ($\alpha_i \in PV^{\bullet,\bullet}$)
\[
S_{int} =\sum_{n\geq 3} \frac{1}{n!} \int_{M}^{PV} \langle u^{k_1} \alpha_1  \otimes \dots \otimes u^{k_n} \alpha_{n} \rangle_0 , \quad \langle u^{k_1} \alpha_1  \otimes \dots \otimes u^{k_n} \alpha_{n} \rangle_0 := \begin{pmatrix}  n-3 \\ k_1 \cdots k_n \end{pmatrix} \, \alpha_1 \wedge \dots \wedge \alpha_n  \, .
\]
Thus the cubic term agrees with BCOV's cubic term. For simplicity, we will often specialize our description to $n=3$ in the remainder of this article.   

\section{Our Worldline Model}\label{sec:models}

The standard sigma-model description of the topological B-model is based on a $2$-dimensional sigma model with a \emph{global} $(2,2)$-supersymmetry. In particular, the (super) reparametrizations are not gauged since the model is defined directly on the moduli space of punctured Riemann surfaces. Furthermore, information on these moduli is lost after the topological twist. To realize Kodaira--Spencer theory as a target space theory for the topological string, the worldline description is based on the isomorphism  
\begin{align}\label{eq:iso}
  \mathcal{T}:  
  C^\infty(U)[\bar\theta^1, \dots ,\bar\theta^n\vert \bar\psi^1, \dots ,\bar\psi^n]\xrightarrow{\sim}  PV^{\bullet,\bullet}(U)\,, \quad U\subset M,
\end{align}
with the former represented as a Fock module of the operator formulation of the  spinning relativistic point particle with global $\mathcal{N}=(2;2)$ supersymmetry, 
 represented in terms of the canonical pairs 
 \begin{align}
     \{\theta,\bar\theta\}= \{\psi,\bar\psi\}=[x,p]=1\,, 
 \end{align}
 where $x$ is a local coordinate on $U\subset M$.\footnote{This realization is quasi-isomorphic to the traditional description of the B-model.} On this module, the supercharges $\bar q=\bar\psi \bar p$ and $q=\theta  p$ represent the Dolbeault differential and divergence on $PV^{\bullet,\bullet}$ with
\begin{align}\label{eq:wlws}
 \{q,\bar q \} = 0= \{q,q\} = \{\bar q,\bar q\}, \qquad q \sim \text{div}, \quad \bar q \sim \bar \partial .   
\end{align}
Following the procedure of the topological string one may then declare $\bar q$ to be the linear BV-differential (implementing holomorphic reparametrizations in the target space) and construct an appropriate complex for $\bar q$. The unconventional feature of this construction is the well-known non-local kinetic term in the action\footnote{Recalling that $\mu$ takes values in $TM\cong M[\bar\theta^i]$ the presence of $\frac{1}{q}$ can be understood from the fact that in $3$ dimensions the natural pairing is between $1$- and $2$-vector fields, rather than two $1$-vector fields.}, as recalled in the previous section but now written in a new parametrization
\begin{align}\label{eq:nlk}
    S=(\mu,\frac{\bar q}{q} \mu) + (\mu, \mu\cdot \mu)\,,
\end{align}
as described in \cite{Bershadsky:1993cx}. This is in contrast with the general expectation \cite{Zwiebach:1992ie} for any string field action where the action for the string field $\psi$ is of the form
\begin{align}
    S=(\psi, Q\psi)+(\psi, [\psi,\psi])+\cdots\,,
\end{align}
where $Q$ implements some linear gauge transformation in target space and $[\cdot,\cdot]$ is a bilinear map on the space of string fields. The origin of this discrepancy is easily traced back to the fact that the worldline/worldsheet model \eqref{eq:wlws} is not a string theory because the worldsheet reparametrizations are not gauged. Therefore, the structure of the target space action is not implied by the BV-structure of the underlying super-moduli space of Riemann surfaces/worldlines. The  motivation of the present paper is twofold: One is to to provide such a connection to the geometry of the relevant super moduli space. This also leads the way to to resolve the non-local kinetic term by importing ideas form the Ramond sector of string theory. However, temporarily ignoring the underlying geometry of super moduli space, we begin in the first subsection by presenting a way to ``localize" the kinetic term in \eqref{eq:nlk} with the help of auxiliary fields. 

\subsection{A local BCOV action with global world line SUSY}
Before establishing a relation to super moduli space we present a simple model to resolve the non-local kinetic term of BCOV from a worldline perspective. 
We choose the ghost degree of an element of $PV^{i,j}$, locally isomorphic to $ C^\infty(U) [\bar\theta^1, \dots,\bar\theta^n\vert \bar\psi^1, \dots ,\bar\psi^n]$, as
\begin{equation}
    gh(v^{(i,j)})= j-i \, .
\end{equation}
Parity is then defined as $gh$ mod $2$. Furthermore, both $\bar q \sim \bar \partial$ and $q \sim \text{div}$ increase this degree. We consider a Calabi--Yau ($n=3$)-fold as our underlying manifold and perform a decomposition in subspaces of definite ghost degree. Take
 \begin{equation}\label{eq:as}
       \tilde a = 
       \mu^{(1,0)}+\mu^{(2,1)} + \mu^{(3,2)},\,
\end{equation}
which forms a multiplet for the non-local and even-parity differential\footnote{We would like to stress that the differential operator has a nilpotent left action.} 
$d=\bar q q^{-1}$ entering in the kinetic term of \eqref{eq:nlk}. Upon acting with $q$ on \eqref{eq:as} this produces a second ``string field"
 \begin{align}
   &
   a = g^{(0,0)} + \mu^{(1,1)} + \mu^{(2,2)}\,,
    \label{st}
\end{align}
which, in addition to the Beltrami differential, also contains the compensator $g^{(0,0)}$. {Note however that $a$ does not form a multiplet for $d$ but we will not need such a structure below.} 
Now we would like to write down a classical action functional $S$ with a kinetic term given by $\bar q$. For this we first  choose an {even} pairing of ghost number $0$
\begin{align}\label{eq:pair22}
\left(a_1, a_2\right)=\int_{CY^3} \mathrm{d}vol_{CY^3} 
\int  < a_1 \, a_2 \,
\Upsilon(\theta^{\wedge 3})\Upsilon(\psi^{\wedge 3})>  \, \quad a_1,a_2 \in C^\infty(U) [\bar\theta^1, \bar\theta^2,\bar\theta^3\vert \bar\psi^1, \bar\psi^2 ,\bar\psi^3]\,.
\end{align}
 In the above, the pointed brackets mean ``normal ordering", i.e.~complete contraction of the $\bar\theta$'s and $\bar\psi$'s. By the isomorphism with polyvectors, this is the same as contraction with a volume form. Moreover, $q=\theta \cdot p$ and $\bar q = \bar\psi \cdot\bar p$ are self-adjoint w.r.t.~the pairing. We then consider the bilinear term 
\begin{equation}
    S_1 = (\tilde a, \bar q a) \,.
    \label{eq:s1first}
\end{equation}
To display this in component fields, we switch to the isomorphic representation in the polyvectors, denoted by pointed bracket $\langle \mathcal{T}- , \mathcal{T}- \rangle \equiv (-,-)$\,:
\[
\langle \mu, \nu\rangle := \int_{CY^3} \Upsilon \wedge  \iota_{\mu\wedge \nu} \Upsilon =  \int_{CY^3}\Upsilon \wedge \iota_\mu (\iota_\nu \Upsilon)\, \quad \mu,\nu \in PV^{\bullet,\bullet}\cong \text{im}\mathcal{T}.
\]
Then $S_1$ becomes\footnote{Due to the isomorphism in \eqref{eq:iso2}, we use the same symbol to denote the component polyvector fields.} 
\begin{equation}\label{eq:s1}
    S_1 = \langle \mu^{(2,1)}, \bar \partial \mu^{(1,1)}\rangle + \langle \mu^{(3,2)}, \bar \partial g^{(0,0)}\rangle + \langle \mu^{(1,0)},\bar \partial\mu^{(2,2)}\rangle \, .
    \end{equation}
We then add the term 
\begin{align}\label{eq:s2}
   S_2 &= \frac 1 2 (\tilde a ,\bar q q \tilde a) = \frac{1}{2}\langle \mu^{(2,1)},\bar\partial\text{div}\mu^{(2,1)} \rangle+\langle \mu^{(3,2)},\bar\partial \text{div}\mu^{(1,0)} \rangle\,,
   \end{align}
together with the cubic interaction 
\begin{equation}
    \frac{1}{3!} (a, a\cdot a) =
    \frac 1 6 \langle \mu^{(1,1)}, \mu^{(1,1)}\wedge \mu^{(1,1)}\rangle  + \langle \mu^{(2,2)},\mu^{(1,1)}\wedge g^{(0,0)}\rangle\,.
    \label{eq:s3}
\end{equation}
Summing up \eqref{eq:s1first}, \eqref{eq:s2} and \eqref{eq:s3} yields Kodaira--Spencer theory. Since this will be extensively analyzed in a related context in \cref{short-multpl}, we will keep the discussion concise here: Half of the Euler-Lagrange equations imply the divergence-less condition for the multiplet $a$. Acting with the div operator on the other half and substituting the solution of the first half of the set yields 
\begin{align}
    \bar\partial g^{(0,0)} + \text{div} (\mu^{(1,1)} g^{(0,0)}) = 0,\\
    \bar\partial \mu^{(1,1)} + \text{div} (\frac 1 2 \mu^{(1,1)} \mu^{(1,1)} + \mu^{(2,2)}g^{(0,0)}) = 0,\\
\bar\partial \mu^{(2,2)} + \text{div} (\mu^{(2,2)} \mu^{(1,1)}) = 0\,.
\end{align}
These are integrability conditions for {the divergenceless} Beltrami differential $\mu^{(1,1)}$, compensator $g^{(0,0)}$ and higher polyvector $\mu^{(2,2)}$. However, this formulation is not yet a string field theory providing the connection to super moduli space. As explained above, the introduction of the auxiliary multiplet $\tilde{a}$ just resolves the non-local kinetic term. The super moduli space and its implications will be addressed in the next subsection. 

\subsection{A model with local world line SUSY}
\label{local20}

In order to obtain a string field theory understanding of the non-locality of the BCOV action we now consider a worldline model with $\mathcal{N}=(2;2)$ supersymmetry where one of the supersymmetries is gauged. A way to describe it is to begin with a supersymmetric version of Baulieu--Singer topological $\sigma$-model \cite{Baulieu:1989rs}. That is, we take a $\bar\partial$ cocycle on an anti-holomorphic curve, $\bar\pi(z)\mathrm{d}\bar z$ and start with the topological action for the parametrized curve  
\begin{align}
    I=\int \bar \pi(z)\cdot \mathrm{d}\bar z\,.
\end{align}
$I$ has a local invariance, $\delta \bar z(t)=\bar \epsilon(t)$. In the BRST formalism this results from the action of an odd vector field $\mathfrak{s}$  
\begin{equation}
\mathfrak{s} \bar z=\bar\psi, \quad \mathfrak{s}\bar \psi=0\,.
\label{BRST1}
\end{equation}
In the canonical formulation, the corresponding BRST-operator is given by $\bar q = \bar \psi \cdot \bar p$, with $\bar p$ represented on functions by $\partial_{\bar z}$.
If we then add the trivial pair $\psi, \bar p$, with \begin{equation}
    \mathfrak{s} \psi=\bar p
    \label{BRST1.1}
\end{equation}
together with the gauge fixing fermion 
\begin{align}
    \Psi= \int \psi \cdot \mathrm{d}\bar z,
\end{align}
one sees that the equivalence holds
\begin{align}
   I\sim I+\mathfrak{s}\Psi=I+ \int \bar p \cdot \mathrm{d}{\bar z}+\psi\cdot\mathrm{d}\bar\psi \,.
\end{align}
To continue we set $\bar\pi\equiv 0$. We then extend this model by a holomorphic sector with a {\it local} odd symmetry on the worldline. The resulting action functional is given by 
\begin{align}\label{eq:Iext}
    I_{\bar\pi=0}=&\int \bar p \cdot \mathrm{d}{\bar z}+\psi\cdot\mathrm{d}\bar\psi +p\cdot(\mathrm{d}z+ \chi\cdot\theta)+\bar\theta\cdot\mathrm{d}\theta+\beta\mathrm{d}\gamma\,.
\end{align}
Here $\chi$ is a worldline gravitino, geometrically a super Beltrami differential on the worldline. The canonical pair $[\beta,\gamma]=1$ represents the Faddeev-Popov super ghosts, of even parity, arising from the gauge fixing of $\chi$. The resulting BRST-functional is invariant under the BRST transformations
\begin{equation}
\mathfrak{s} z = \gamma \theta , \quad \mathfrak{s} \bar\theta = \gamma p_z , \quad \mathfrak{s} \beta = \theta\cdot p \,
\label{BRST2}
\end{equation}
with the remaining transformations being trivial. In the operator formulation, after gauge fixing $\chi$, we are left with two BRST differentials represented by the operators
\begin{align}\label{eq:diff}
    \bar q=&\bar\psi\cdot\bar p\, , \\ 
    \gamma q =&\gamma p\cdot\theta\,. \label{yq}
\end{align}
{\bf Remark:} Let us emphasize that the algebra generated by $q$ does not close into a translation and is thus not a supersymmetry but rather a nilpotent symmetry on $\mathbb{R}^{0|1}$. We can promote this to a $\mathcal{N}=2$ SUSY by adding the terms $\bar p\cdot\bar\chi(\bar\theta)-\frac{e}{2}\bar p\cdot p$ to the worldline action. We will comment on this point in \cref{BRST-22}. 

\noindent{\bf Remark:} The action \eqref{BRST2} is not the dimensional reduction of the (global) $\mathcal{N}=(1,1)$ sigma model which gives rise to the topological string after the ``topological twist" \cite{Marino:2004eq}. The latter is defined directly on the moduli space and therefore lacks the $\beta$- and $\gamma$ superghosts which play an important role in our construction below. In particular, the $\gamma$-ghost will provide a geometric interpretation of the equivariant parameter $u$ in section \ref{sec:rev}. Furthermore, the action \eqref{BRST2} is not topological in the sense that it still depends on the odd moduli that parametrizes the gravitino $\chi$ modulo super reparametrizations. The path-integral evaluation with the locally symmetric action \eqref{eq:Iext} involves integration over a parity odd subbundle of super moduli space $\mathcal{M}=\mathcal{J}/D$ of $\chi$'s modulo globally defined odd reparamerizations (see e.g. \cite{Witten:2012bh,Cremonini:2025eds} for details). In string theory, integration over this subbundle gives rise to {\emph{picture changing}}, which we recall now.

 

\subsubsection*{Summary on pictures}
In a (differential) superLie algebra $\mathfrak{g}$ some of the generators have odd parity therefore, when doing a degree shift as prescribed in BRST and BV, these odd generators become even. So  $(\mathfrak{g}\oplus\mathfrak{g^*})[1]$ has  the structure of a direct sum of Clifford and Weyl algebra, with the latter generated by the superghosts $\gamma$ and $\beta$. Moreover the Chevalley--Eilenberg complex is unbounded, so standard integration is not defined. For this reason one introduces distributional valued forms \cite{Belopolsky:1997bg}. This representation sits in a different \textbf{picture} than the standard representation of Lie algebra forms, $\Lambda^\bullet \mathfrak{g}$.
Hence with \textbf{picture number} it is meant \emph{the number of shifted odd generators (now of even parity) whose highest weight states are Dirac deltas}, $\delta(\gamma)$. 
Correspondingly, at each picture there arises a cochain complex, graded by the differential for the superLie algebra. \emph{Integral forms} sit in the cochain complex with maximal picture number, while \emph{super forms} have zero picture and the intermediate complexes host \emph{pseudoforms}.
A \textbf{picture changing operator} is a cochain map, that relates cochains with different pictures. We denote the cochains with picture $p$ as $C^p$. In our context   $\bar q$ is the picture-preserving differential for the cochain complex. Then 
\[
X^{(p)}: C^p \to C^{p\pm1}, \quad X^{(p)} \bar q = \bar q X^{(p)}
\]
depending on whether $X^{(p)}$ lowers or increases the picture. Furthermore, the picture changing operators are cohomology classes, $X^{(p)} \neq [\bar q,-]$. 

In our representation of Kodaira-Spencer gravity, picture changing is realized as a parity even operator
\[
X = \delta(\beta) \, \theta \cdot p\,,
\]
giving rise to the divergence operator, ubiquitous in BCOV. So, in our model, super moduli space is the geometric origin of this operator.

\subsubsection*{Kodaira--Spencer theory from the worldline}
After this preparation, we are ready to formulate Kodaira--Spencer theory on a suitable module for the operator algebra just described. Given an open subset $U$ of a compact $CY^3$, the representation space will be a submodule of 
\begin{equation}
C^\infty(U)[\bar\theta^1, \bar\theta^2 ,\bar\theta^3\vert \bar\psi^1, \bar\psi^2 ,\bar\psi^3] \otimes (\mathbb{C}((\gamma)) \oplus \mathbb{C} [[\partial_\gamma]])\,,
\end{equation}
where $\partial_\gamma$ is understood to act on the Dirac delta distribution, $\delta(\gamma)$. The situation just described can also be called the \emph{small Hilbert space}, in opposition to the \emph{large Hilbert space} that will appear later. The module is graded according to the following convention:
\begin{align}
   \text{gh}(\bar\theta)= \text{gh}(\gamma) = -\text{gh}(\bar\psi) =  -\text{gh}(\delta(\gamma))= 1 ,
    \label{ghostdeg}
\end{align}
and we will write $\bigoplus_k E^k$ for the module.
We endow the graded module with an odd, ghost degree $1$ pairing 
\begin{align}\label{eq:pairns}
(-,-):&\, E^{k_1} \times E^{k_2} \to \mathbb{C}[k_1+k_2+1],\nonumber\\
\left(a,b\right)=&\int_{CY^3} \mathrm{d}vol_{CY^3} 
\int \, \;\mathrm{d}\gamma\; < a \, b\, 
\Upsilon(\theta^{\wedge 3})\Upsilon(\psi^{\wedge3})>\,,
\end{align}
where $<\cdots >$ stands for the complete contraction of $\theta$'s and  $\bar \theta$'s and similarly for $\bar\psi$ (see \cite{Cremonini:2025eds}). The dependence on $\bar\theta$ and $\bar \psi$ is removed thus leaving us with a function on the Calabi--Yau. 
The integral over $\gamma$ is odd and implies that our pairing has ``picture 1". It is thus clear why, alongside with polynomials in $\gamma$, we have to include a dual multiplet which is distributional (Dirac delta) in $\gamma$, rendering the parity of the integrand odd.

In fact, as the pairing together with a kinetic operator makes up the two-point functions, the relevant worldline has just two punctures, so the moduli space is just a point \cite{Cremonini:2025eds}. Then  $\gamma$ and $\delta(\gamma)$ are the Faddeev--Popov ghosts for setting to zero the global odd transformations in \eqref{eq:Iext}.

What submodule do we want? The fields we keep are summarized in the following table, that shows the ghost number and parity of the multiplets, down to their component fields:
  \begin{center}
                \begin{tabular}{c|c|c|c||c|c|c|c||c|c|c|c}
                    \multicolumn{4}{c||}{multiplet} & \multicolumn{4}{c||}{parity} & \multicolumn{4}{c}{ghost number}\\
                    \hline \hline
                     \multicolumn{4}{c||}{$a^{(0)}$} &  \multicolumn{4}{c||}{even} & \multicolumn{4}{c}{0} \\
                     \hline
                     $\mu^{(1,0)}$ & $\mu^{(1,1)}$ & $\mu^{*(1,2)} $& $ \mu^{*(1,3)}$  & odd & even& {odd} & {even}  &-1 & 0 & 1 & 2\\
                     \hline \hline
                     \multicolumn{4}{c||}{$a^{(-1)}$} &  \multicolumn{4}{c||}{even} & \multicolumn{4}{c}{0}\\
                     \hline
                     $\mu^{(2,0)} \delta(\gamma) $&$ \mu^{(2,1)}\delta(\gamma)$ &$\mu^{*(2,2)}\delta(\gamma) $&$ \mu^{*(2,3)} \delta(\gamma)$ & odd & even& {odd} & {even} &-1 & 0&  1 & 2\\
                     \hline
                \end{tabular}
                \captionof{table}{} 
        \end{center}
So we wish to restrict our considerations to polyvectors of multivector degree $1$ or $2$. Some explanation on why these multiplets and what could go wrong with different choices is due. First of all, $a^{(0)}$ is a minimal set of fields {in ghost number $0$} containing the Beltrami differential:
\begin{align}\label{eq:a0}
    a^{(0)}&= \mu^a\bar\theta_a+\mu^b_a\bar\psi^a\bar\theta_b++\mu^b_{ac}\bar\psi^a\bar\psi^c\bar\theta_b+\mu^b_{acd}\bar\psi^a\bar\psi^c\bar\psi^d\bar\theta_b\equiv{\mu^{(1,0)}}+{\mu^{(1,1)}}   +{\mu^{*(1,2)}}+{\mu^{*(1,3)}}\,, 
\end{align}
where $\mu^{(1,0)}$ is a gauge symmetry for the Beltrami differential $\mu^{(1,1)}$ with the differential $ \bar q$ of degree $-1$. 
The remaining terms complete a multiplet for $\bar q$, and will be interpreted as antifields but for another set of component fields. 

By virtue of our pairing \eqref{eq:pairns} the dual to $a^{(0)}$ is determined to be the picture $-1$ multiplet: 
\begin{align}\label{eq:a0*}
    a^{(-1)}&= {\mu^{(2,0)}}\delta(\gamma)+{\mu^{(2,1)}}\delta(\gamma) + {\mu^{*(2,2)}}  \delta(\gamma) +{{\mu^{*(2,3)}}}\delta(\gamma)\,.
\end{align}
Another option would be to have the Beltrami differential in a picture $-1$ multiplet, based on the $\delta(\gamma)$. In that case an interaction term would make up an integral form on $\mathcal{M}$ instead, e.g. \cite{Cremonini:2025eds}.

We thus arrange the component fields into two cochain complexes related by the non-invertible cochain map $X$, a picture raising operator that commutes with $\bar q$
\begin{align}
    X=\delta(\beta) q\,,\qquad \text{gh} X= 0\,,
    \label{ghostofX}
\end{align}
\begin{equation}\label{q:blockseq}
\begin{tikzcd}
 {{\mu^{(1,0)}}}  \arrow{r}{\bar{ q}} & {{\mu^{(1,1)}}} \arrow{r}{\bar{ q}}  &  {{\mu^{*(1,2)}}} \arrow{r}{\bar q}  & {{\mu^{*(1,3)}}}\\%
 \delta(\gamma){\mu^{(2,0)}} \arrow{u}{ X} \arrow{r}{\bar{ q}}  & \delta(\gamma) {{\mu^{(2,1)}}}  \arrow{u}{ X}\arrow{r}{\bar{ q}}& \delta(\gamma) {{\mu^{*(2,2)}} }\arrow{u}{ X}\arrow{r}{\bar{ q}}  & \delta(\gamma) { {\mu^{*(2,3)}}}\arrow{u}{ X}\,.
\end{tikzcd}\,   
\end{equation}
Restriction of $(-,-)$ to this total complex yields a non-degenerate pairing by construction. The operators $\bar q$ and $X$ respect the following conditions with the pairing:
\begin{align}
    (\bar q a, b ) = -(-1)^{\vert a \vert}(a, \bar q b) , \quad (X a, b ) = (-1)^{\vert a \vert} (a, X b) 
\end{align}
The parity $\vert - \vert$ is according to Table $1$ but here we only deal with even multiplets.

Below we will often encounter the  pairing where $\gamma$ is integrated. This is natural the pairing for polyvectors, with pointed bracket $\langle -,-\rangle$. Anyway, the $\mathbb{Z}_2$ degree for the components is still as in Table $1$.  Later we will  promote the pairing $\langle -,-\rangle$, shifted by $(-1)$, to an odd Poisson bracket. 

Diagram \eqref{q:blockseq} already gives indications why the compensator field $g$ reviewed in Appendix \ref{App} does not show up in this construction. It would need to have picture $+1$ which is  impossible (since it would correspond to a super form of negative form degree). Furthermore, in the operator formalism the action of $X$ on the picture zero multiplet does not vanish unless the multiplet is divergenceless. From this observation we may already suspect that gauge invariance in {the small Hilbert space} requires divergencelessness, as we indeed confirm below. 

The double complex \eqref{q:blockseq} is just what we need to produce a local BV-quadratic action for the Beltrami differential. Indeed, in BCOV the quadratic term is non-local: here that problem is naturally mapped to the problem of pictures in the Ramond sector of open super string theory \cite{Konopka:2016grr}. There, as well as here the naive, natural kinetic term of the BV-action
\begin{align}\label{eq:S2att}
 S_2=   (a^{(0)}, \bar q a^{(0)})
\end{align}
is not admissible since the picture does not add up to $0$, i.e.~the integration over $\gamma$ is not well defined\footnote{If we were to insert a $\delta(\gamma)$ in the pairing \eqref{eq:pair}, then \eqref{eq:S2att} would be well defined and would instead result in the appropriate kinetic term for Chern--Simons theory.}. This is precisely what happens in the Ramond sector of string theory. In that theory there is a solution to this problem by introducing an auxiliary field with picture shifted by $-1$ \cite{Sen:2019qit}. This then suggests a solution in our case as well, and justifies why we have introduced $a^{(-1)}$. 

Thanks to that, now we have a well defined kinetic term. Super moduli space motivates $X$ but not the form of the action. The following expression is motivated by variational calculus: its field equations have to be the condition of $\bar q$-closure. 
\begin{align}
    S_2&=\frac{1}{2}(a^{(-1)},\bar q X a^{(-1)}) + (a^{(-1)},\bar q  a^{(0)})\label{eq:freepics0}\\
    &=\frac{1}{2}\langle \mu^{(2,1)},\bar\partial\text{div}\mu^{(2,1)} +2 \bar\partial \mu^{(1,1)}\rangle + \langle \mu^{(2,0)}, \bar\partial \text{div}\mu^{*(2,2)} +\bar\partial \mu^{*(1,2)} \rangle +  \langle \mu^{*(2,2)},\bar\partial\mu^{(1,0)} \rangle\,,
    \label{eq:freepics}
    \end{align}
where, in the second line, we integrated the modulus $\gamma$ and used the isomorphic representation in the polyvectors as before. Note that the functional $S_2$ is even. We shall see that the fields equations capture Dolbeault cohomology, where $\mu^{(1,1)}$ and $\mu^{(1,0)}$ are also div-closed. Note that, {as already anticipated by the diagram \eqref{q:blockseq}}, the field $g$ is missing. 
The fact that the compensator $g$ is absent, while not an inconsistency, is a shortcoming that will be addressed in the large Hilbert space description \cref{sec:lhs}. Anyway, the div operator appears naturally in this description in the form of picture changing as a consequence of working on super moduli space. An alternative way to match the picture without introducing auxiliary fields would be to insert the inverse $X^{-1}$ of the picture raising operator into  \eqref{eq:S2att}. However, the inverse is not well defined since $X$ has a non-vanishing (co)-kernel. This is just the same problem of the $\text{div}^{-1}$ operator in the BCOV-kinetic term, which we propose to address in this paper.

Now consider the bracket on the component fields of the polyvectors induced by the non-degenerate pairing $\langle-,-\rangle$, shifted by $-1$, that pairs odd fields to even fields: 
\begin{equation}
        \{\mu^{i_1,j_1}(z_1), \mu^{i_2,j_2}(z_2)\} = \delta^{i_1+i_2,3}\delta^{j_1+j_2,3} \delta(z_1-z_2)\Upsilon(\bar\theta^3)\Upsilon(\bar\psi^3)\,.
        \label{Poisson-ndg}
    \end{equation}
This extends to an {odd} Poisson bracket, so that the 
quadratic action \eqref{eq:freepics} is a genuine BV theory that solves the classical master equation as a consequence of nilpotency of the differential:
\begin{equation}
\{S_2,S_2\}=0.
\end{equation}
This feature should not come as a surprise: in fact, the $L_\infty$ structure under inspection is simply a differential graded algebra, i.e.~polyvector fields endowed with the $\bar \partial$ differential.
The action makes the role of $\mu^{(1,0)}$ and $\mu^{(2,0)}$ as gauge parameters for  $\mu^{(1,1)}$ and $\mu^{(2,1)}$ manifest. Geometrically, $\mu^{(1,0)}$ represents holomorphic reparametrizations.

\subsection{Interacting theory and its BV formulation}
\label{shs-BVint}
Let us now turn to the cubic interaction. For the $3$-punctured line the moduli space $\mathcal{M}\simeq \mathbb{R}^{0|1}$ is $(0|1)$-dimensional \cite{Cremonini:2025eds}. {One way to eliminate the} dependence of this contribution on $\mathcal{M}$ {is to insert a} Poincaré  dual $Y=\eta\delta(\mathrm{d}\eta)$ in the path integral, thus eliminating all dependence on the odd coordinates (see, e.g. \cite{Cremonini:2025eds}). Then the $3$-point correlator representing the cubic term should produce a function{, evaluated at $0$,} rather than a pseudoform on $\mathcal{M}$. This can be achieved with an insertion in picture zero. In addition, we insert a $\delta(\gamma)$ (anywhere on the line) to provide the Jacobian for isolating the global odd transformation\footnote{Alternatively, we could have one insertion in picture $-1$ and two insertions in picture $0$. However, consulting \eqref{q:blockseq} we can see that there is not a non-vanishing contraction of that kind. This is implied by counting of the vector field degree.}. 
The goal of this is to implement a DGLA structure where the $2$-product will come out to be the Schouten--Nijenhuis bracket of multivector fields. 
Adding the cubic interaction to $S_{2}$ we end up with
\begin{align}\label{eq:SBVs}
 S_{SHS}=&\frac{1}{2}(a^{(-1)},\bar q X a^{(-1)}) + (a^{(-1)}, \bar q  a^{(0)}) +\frac{1}{3!} (a^{(0)}, \delta(\gamma) a^{(0)}\wedge a^{(0)})
 \\
 =& \frac 1 2 \langle \mu^{(2,1)},\bar\partial\text{div}\mu^{(2,1)} \rangle+\langle \mu^{(2,1)},\bar\partial\mu^{(1,1)} \rangle + \langle \mu^{(2,0)}, \bar\partial \text{div}\mu^{*(2,2)} +\bar\partial \mu^{*(1,2)} \rangle +  \langle \mu^{*(2,2)},\bar\partial\mu^{(1,0)} \rangle \nonumber\\
  & +\frac{1}{3!}\langle  \mu^{(1,1)},\mu^{(1,1)}\wedge \mu^{(1,1)}\rangle + \langle \mu^{(1,0)} , \mu^{(1,1)} \wedge \mu^{*(1,2)} + \frac 1 2 \mu^{(1,0)} \wedge \mu^{*(1,3)} \rangle \,.\label{reduced_nogamma}
\end{align}
{\textbf{Remarks}: Despite its suggestive formula, this functional cannot possibly solve the classical master equation with the bracket in \eqref{Poisson-ndg} since the divergence operator is not a derivation of the wedge product. Therefore, there is no $L_2$ structure as of now, but we will see that it will appear when reducing down to the locus of some of the field equations.}

If we remove $\mu^{(1,0)}$ and $\mu^{*(2,2)}$ then the action is again a BV action with a local kinetic term. It solves the CME with the bracket \eqref{Poisson-ndg} because of nilpotency of the differential, and the interaction does not enter the computation.  However, $S_{SHS}$ is not invariant under  the expected gauge symmetry for the Beltrami differential. We believe that this shortcoming can be understood geometrically: Insertion of picture $0$ operators in the interaction term as described above does produce a function as required, but it does not guarantee that this function does not depend on the odd coordinate $\eta$ on super moduli space, $\mathbb{R}^{0\vert1}$. Another way of seeing this is the missing picture changing $X$ in the cubic term. The latter provides the chain map 
\begin{align}
    X\circ\bar q=\int_{\mathcal{M}} \mathrm{d}\circ X
\end{align}
between the BRST differential on the module and the de Rham differential on super moduli space (e.g.~\cite{Witten:2012bh,Cremonini:2025eds}) which, in turn,  relates gauge-invariance to the absence of boundaries on super moduli space. This observation then also shows the way to restore gauge invariance. If the  $a^{(0)}$-insertions in the cubic term are in the image of $X$, the chain map property holds. In fact, it is sufficient for $a^{(0)}\in \ker X$. It turns out that we can implement this substitution directly as a symplectic reduction at the level of the action: For that, we note that the E-L equations for the fields $\mu^{(2,1)}$ and $\mu^{(2,2)}$ are
\begin{align}\label{eq:KSsmall}
    \bar q (\text{div} \mu^{(2,0)} + \mu^{(1,0)}) = 0\,,\\
    \bar q (\text{div}  \mu^{(2,1)} + \mu^{(1,1)}) =0\,. \label{forBeltrami}
\end{align}
If $\mu^{(1,1)}$ has no projection on the cohomology of $q$, the second equation can be solved to express $\mu^{(1,1)}=-\text{div} \mu^{(2,1)} $, implying, in particular, that $\mu^{(1,1)}$ is divergence free. If, on the other hand, $\mu^{(1,1)}\in H_{\bar\partial}(PV^{})$ then, for a compact CY, 
it follows from the $\partial\bar\partial$-lemma that $\mu^{(1,1)}$ can be assumed to be divergence-free (there is a quasi-isomorphism between $H_{\bar\partial}(PV)$ and $H_{\bar\partial}(\ker \, \text{div})$, see \cite[lemma 6.1]{Barannikov:1997dwc} and for the proof see \cite[section 6]{DDlemma-Deligne75}). The same considerations apply to $\mu^{(1,0)}$. Using the above equation \eqref{forBeltrami} on the $\mu^{(1,1)}$-field equation hit by $\text{div}$, we get 
\begin{align}
     \bar q \mu^{(1,1)} +\frac{1}{2}\text{div}(\mu^{(1,1)} \wedge \mu^{(1,1)})&=0\,,
\end{align}
which is equivalent to the KS-equation for $\mu^{(1,1)}$ or, in other words to the integrability condition of the Beltrami differential. Here, the Bogomolov--Tian--Todorov lemma \cite{Tian1987SmoothnessOT} is used (see  Appendix \eqref{Tian-version}). 

\noindent \textbf{Remark:}
One may wonder if the conditions \cref{eq:KSsmall,forBeltrami} could be formulated by means of a gauge fixing fermion. That is the case: Alternatively to the projection, we can do a symplectic reduction with the help of a gauge fixing fermion
\begin{align}
    \Psi=b^{(2,1)}\bar q (q  \mu^{(2,1)} + \mu^{(1,1)})+b^{(2,3)} (q  \mu^{(2,0)} + \mu^{(1,0)})
\end{align}
where $(b^{(2,1)}\stackrel{\bar q}{\to} b^{(2,2)}\stackrel{\bar q}{\to} b^{(2,3)})$ is an additional \emph{trivial multiplet}.

Let us follow what happens for the action \eqref{reduced_nogamma}, after going to the locus of the field equations \cref{eq:KSsmall,forBeltrami}: We express the \emph{divergence-free} $\mu^{(1,1)}$ in terms of $\mu^{(2,1)} $ and the harmonic part $h$. Recall that on a compact CY it holds that  $\bar\partial h = 0$. The same considerations apply to $\mu^{(1,0)}$ which we set to $\mu^{(1,0)}= -\text{div} \mu^{(2,0)}$. Then we have
\begin{align}
S_{red}[\mu]= & -\frac 1 2 \langle \mu^{(2,1)},\bar\partial \text{div}\mu^{(2,1)}\rangle -\frac{1}{3!} \langle  \text{div}\mu^{(2,1)}+h, (\text{div} \mu^{(2,1)}+h) \wedge   (\text{div} \mu^{(2,1)}+h)\rangle \notag \\
 & -\langle  \mu^{*(1,2)},\bar\partial \mu^{(2,0)} + {{\text{div} \mu^{(2,0)} \wedge  (\text{div} \mu^{(2,1)} + h)}} \rangle -\frac{1}{2} {{\langle  {\mu}^{*(1,3)}, \text{div} \mu^{(2,0)} \wedge \text{div} \mu^{(2,0)}\rangle}} \,.
 \label{BV-imdiv}
\end{align}
The action $S_{red}$ is formulated on a symplectic subspace of the original space, i.e.~with a non-degenerate symplectic form which thus gives rise to a non-degenerate BV-bracket. However, the action \eqref{BV-imdiv} is not a solution of the BV-master equation unless the antifields ${\mu}^{*(1,2)} $ and ${\mu}^{*(1,3)}$ are set to be div-exact. {We will do so,} after evaluating the BV-bracket. Composing with such projection to the div-exact subspace, implies that the result is a \emph{non-invertible} $(-1)$-shifted Poisson bracket. On functionals $S,F$ on the symplectic reduced space, it is given by
\begin{equation*}
           \{S,F\} = \left.\left(\frac{\delta S}{\delta \mu^{2,j_1}} \frac{\delta F}{\delta \mu^{*1,j_2}} - \frac{\delta S}{\delta \mu^{*1,j_2}} \frac{\delta F}{\delta \mu^{2,j_1}} \right) \right|_{\overset{\mu^* \in \text{div} PV }{j_1<2\leq j_2}} \; 
        \delta^{j_1+j_2,3} \delta(z-z') \Upsilon(\bar\theta^3)\Upsilon(\bar\psi^3)\, .
        \label{symplform}
    \end{equation*}

Indeed, the action functional is invariant under the BV-transformations $\delta a = \{S, a\}$:
\begin{align}
    \delta {\mu}^{(2,1)} &= \bar\partial \mu^{(2,0)} + \text{div}\mu^{(2,0)} \wedge (\text{div} \mu^{(2,1)} + h), \notag \\ \delta \mu^{*(1,2)} &= - \bar\partial \text{div} {\mu}^{(2,1)} -\frac 1 2  \text{div}((\text{div}  \mu^{(2,1)} + h) \wedge (\text{div} {\mu}^{(2,1)} + h)) + {{\text{div}(\mu^{*(1,2)} \wedge \text{div} \mu^{(2,0)})}}\vert_{\mu^{*(1,2)}\in\text{div}PV^{(2,2)}}, \notag \\
    {{\delta \mu^{(2,0)}}} &= {{\frac 1 2  \text{div} \mu^{(2,0)} \wedge \text{div} \mu^{(2,0)} }},\notag \\
    {{\delta  \mu^{*(1,3)}}} &= {{\bar\partial \mu^{*(1,2)} + \text{div}( \mu^{*(1,2)} \wedge (\text{div}  \mu^{(2,1)} +h) ) + \text{div}( \mu^{*(1,3)} \wedge \text{div} \mu^{(2,0)})\vert_{\mu^{*(1,\bullet)}\in\text{div}PV^{(2,\bullet)}}}},
\end{align}
as can be checked using integration by parts, $\bar\partial^2=0$, the derivation property of $\bar\partial$ w.r.t.~the alternating product $\wedge$, the relation $\int \text{div} a \wedge b = \text{bdry} -(-1)^{\vert a \vert} \int a \wedge \text{div} b$ and that $\text{div}$ yields the Schouten bracket on a product of \emph{div-free polyvectors}, which satisfies Jacobi identity. For the same reasons, \eqref{BV-imdiv} satisfies the classical BV-master equation. 
    
Then 
\begin{align}
-\{S_{red},S_{red}\} =  &  \,\left\langle \bar\partial \text{div} {\mu}^{(2,1)}+  \frac 1 2 \text{div}((\text{div}\mu^{(2,1)}+h) \wedge  (\text{div} \mu^{(2,1)}+h))  , \bar\partial \mu^{(2,0)} + \text{div}  \mu^{(2,0)} \wedge (\text{div}  \mu^{(2,1)} +h)\right\rangle
\notag \\
& +  \langle  \text{div} (\mu^{*(1,2)} \wedge \text{div} \mu^{(2,0)}), \bar\partial \mu^{(2,0)} + \text{div}  \mu^{(2,0)} \wedge (\text{div}  \mu^{(2,1)} +h) \rangle \notag \\
& + \frac{1}{2} \langle  \text{div} ( \mu^{*(1,3)} \wedge \text{div} \mu^{(2,0)}), \text{div} \mu^{(2,0)} \wedge \text{div} \mu^{(2,0)} \rangle \notag \\
&+\frac 1 2 \langle \bar\partial \mu^{*(1,2)} + \text{div}((\text{div}  \mu^{(2,1)} +h)\wedge \mu^{*(1,2)}),  \text{div} \mu^{(2,0)} \wedge \text{div} \mu^{(2,0)}\rangle
\end{align}
is zero after the aforementioned algebraic massaging. Note that the gauge symmetry of the Beltrami can be obtained by doing these manipulations: 
\begin{equation}\label{eq:dm11}
\delta \mu^{(1,1)} = q \delta\mu^{(2,1)} = -\bar\partial \lambda^{(1,0)} + [\lambda^{(1,0)}, \mu^{(1,1)}] 
\end{equation}
which is equivalent to that from evaluation with the BCOV's bracket. Another expression for BCOV's bracket of \cite{Bershadsky:1993cx}, more convenient for the present situation, is:
    \begin{equation*}
           \{S,F\}_{BCOV} = \left.\left(\left(\text{div}\frac{\delta S}{\delta \mu^{i_1,j_1}} \right)\frac{\delta F}{\delta \mu^{*i_2,j_2}} - \frac{\delta S}{\delta \mu^{*i_2,j_2}} \left(\text{div}\frac{\delta F}{\delta \mu^{i_1,j_1}} \right) \right)\right|_{{i_2+j_2>2}} \!\!\!\!\!\!\!\!\!\!\!\!\!\!\!\delta^{i_1+i_2,2}
        \delta^{j_1+j_2,3} \delta(z-z') \Upsilon(\bar\theta^3)\Upsilon(\bar\psi^3)
    \end{equation*}
Recalling that the space of fields in BCOV is $\ker \text{div} \subset PV^{i,j}$ with $i\leq 2$. Then,
\begin{equation}
    \delta \mu^{(1,1)} = \{S_{red},\mu^{(1,1)}\}_{BCOV}.
\end{equation}
coincides with \eqref{eq:dm11}.

\section{Large Hilbert space}
\label{sec:lhs}

A shortcoming of the formulation in the last section is that the compensator, $g$ for the holomorphic volume form in  \cref{eq:ELCostello2} is not included in the BV-multiplet. In this section we attempt to  get around this by formulating the theory in the \emph{large Hilbert space} for the $\gamma$-ghost, which is achieved by allowing for negative powers of $\gamma$ in the multiplet instead of pictures. Generally speaking, going to the large Hilbert space has proven fruitful in string field theory literature, see for instance \cite{Berkovits:2012np}. Here the large Hilbert space is attractive since it allows to combine multiplets that had different pictures into a single multiplet. On the other hand, in the large Hilbert space we lose the connection to super moduli space. A consequence of this is that there is not a canonical kinetic term in the target space action, which in turn determines the differential for the complex of target space fields. Below we will explore different choices for the latter.

\subsection{A long multiplet}
In the large Hilbert space, what previously was the picture changing operator is now represented by 
\[X=\gamma\theta^a  p_a
\]
and defines a second  differential in addition to $\bar q$. Akin to Costello and Li \cite{Costello:2012cy}, we consider the {$2$-shifted} module {of ``decorated" polyvectors}
\[
 V= C^\infty(U)[\bar\theta^1, \bar\theta^2 ,\bar\theta^3\vert \bar\psi^1, \bar\psi^2 ,\bar\psi^3] \otimes \mathbb{C}[[\gamma]] \, [2]\,,
\]
{which has infinite multiplicity as a power series in $\gamma$. A finite dimensional submodule,} which combines $a^{(0)}$ and $a^{(-1)}$ of section \ref{local20} into a double complex of fields is 
\begin{equation}\label{q:blockseq2}
\begin{tikzcd}
 {\color{black}{\gamma g^{(0,0)}}} \arrow{r}{\bar{ q}} & {\color{black}{\gamma g^{*(0,1)}}} \arrow{r}{\bar{ q}} & {\color{black}{\gamma g^{*(0,2)}}} \arrow{r}{\bar{ q}} & {\color{black}{\gamma  g^{*(0,3)}}}\\
 {{\mu^{(1,0)}}}\arrow{u}{ X}  \arrow{r}{\bar{ q}} & {{\mu^{(1,1)}}}\arrow{u}{ X} \arrow{r}{\bar{ q}}  &  {{\mu^{*(1,2)}}}\arrow{u}{ X} \arrow{r}{\bar q}  & {{\mu^{*(1,3)}}}\arrow{u}{ X}\\%
 \frac{1}{\gamma}{\mu^{(2,0)}} \arrow{u}{ X} \arrow{r}{\bar{ q}}  & \frac{1}{\gamma} {{\mu^{(2,1)}}}  \arrow{u}{ X}\arrow{r}{\bar{ q}}& \frac{1}{\gamma} {{\mu^{*(2,2)}} }\arrow{u}{ X}\arrow{r}{\bar{ q}}  & \frac{1}{\gamma} { {\mu^{*(2,3)}}}\arrow{u}{ X}\\
 \frac{1}{\gamma^2}{g^{(3,0)}} \arrow{u}{ X} \arrow{r}{\bar{ q}}  & \frac{1}{\gamma^2} {{g^{(3,1)}}}  \arrow{u}{ X}\arrow{r}{\bar{ q}}& \frac{1}{\gamma^2} {{g^{(3,2)}} }\arrow{u}{ X}\arrow{r}{\bar{ q}}  & \frac{1}{\gamma^2} { {g^{*(3,3)}}}\arrow{u}{ X} \, ,
\end{tikzcd}\,   
\end{equation}
where we now include states in positive "picture" since there is no geometrical obstruction to this in the large Hilbert space. 

On the matter of the ghost degrees assignment, we follow \cite{Costello:2012cy} and opt for:
\begin{equation}
\text{gh}(\gamma^k PV^{i,j}) =2-(i+j+2k),
\label{gh_CosLi}
\end{equation}
so that fields with homogeneous ghost degree lay in the  diagonals pointing downwards in the double complex.

Furthermore, we need to reconsider the definition of the symplectic form in the large Hilbert space. For this we will represent the delta function as a residue (see e.g.~\cite[chapter 3]{Griffiths:1994prl}) 
\begin{align}\label{eq:pairnL}
\left(a,b\right)=\frac{1}{2\pi i}\int_{CY^3} \mathrm{d}vol_{CY^3}\oint\mathrm{d}\gamma\;< a  \, b \, \Upsilon(\psi^{\wedge 3})\Upsilon(\theta^{\wedge 3})> .
\end{align}
In our paring, the sign convention differs form that of the sesquilinear form in \cite{Costello:2012cy}. We pair $\gamma^k$ with $(\gamma)^{-1-k}$ whereas the latter pairs  $\gamma^k$ with $(-\gamma)^{-1-k}$. The  geometric interpretation of the contour integral in \eqref{eq:pairnL} can be obtained by noticing that there is no reality condition for $\gamma$ in \eqref{eq:diff}. Notice that this pairing is even. 
 
{\bf Remark:} The pairing as defined in \eqref{eq:pairnL} introduces a tautological $S^1$-action: $\gamma\mapsto e^{i\alpha} \gamma$. The latter can be complemented by  $\theta\mapsto e^{-i\alpha} \theta$ to represent the canonical $R$-symmetry present in representations of supersymmetry algebras such as \eqref{eq:Iext}. In our setting the circle operator $D$ of \cite{Costello:2019jsy} is realized as $D=q=\theta^ap_a\sim \text{div}$, with $\rho=[\bar q,D]=0$ on the representation space $V$, in agreement with \cite{Costello:2019jsy}. This also gives an interpretation of the formal parameter $u$ there by identifying $u$ with the super ghost $\gamma$ with $Q=\bar q+\gamma \theta^ap_a =\bar q + X$ reproducing the the equivariant BRST differential of \cite{Costello:2019jsy}.
 
It turns out that the double complex  \eqref{q:blockseq2} can also be be organized as a $Q$-complex, $Q= \bar q + X$
\begin{align}\label{eq:Q-plex}
&\left(\frac{1}{\gamma^2} g^{(3,0)}\right)\stackrel{Q}{\to}\left(\frac{1}{\gamma^2} g^{(3,1)}+\frac{1}{\gamma}{\mu^{(2,0)}}\right)\stackrel{Q}{\to}\nonumber\\
    &\left(\frac{1}{\gamma^2} g^{(3,2)}+\frac{1}{\gamma}{\mu^{(2,1)}}+{\mu^{(1,0)}}\right)\stackrel{Q}{\to}\left(\frac{1}{\gamma^2} g^{*(3,3)}+\frac{1}{\gamma}{\mu^{*(2,2)}}+{\mu^{(1,1)}}+\gamma g^{(0,0)}\right)\nonumber\\&\stackrel{Q}{\to}\left(\frac{1}{\gamma} {\mu^{*(2,3)}}+{\mu^{*(1,2)}}+\gamma g^{*(0,1)}\right)
    \stackrel{Q}{\to}\left( {\mu^{*(1,3)}}+\gamma g^{*(0,2)}\right) \stackrel{Q}{\to}\left( \gamma g^{*(0,3)}\right)
 \end{align}
Each bracket in this complex forms itself a complex w.r.t.~the non-local and even-parity differential $\mathrm{d}=\bar q X^{(-1)}$, which is, however,  well defined only in im$X$. In our convention of section \ref{local20}, \cref{ghostdeg}, the brackets in \eqref{eq:Q-plex} do not have homogeneous ghost number but they do in the convention \eqref{gh_CosLi} of \cite{Costello:2019jsy}. We can thus assemble the entire $Q$-complex into a parity even, ghost number zero multiplet $\mathcal{A}$. Then, the parity odd quadratic action functional  
 \begin{align}
     (\mathcal{A},Q\mathcal{A})=& \,\langle g^{(3,2)},\bar q g^{(0,0)}+q \mu^{(1,1)}\rangle+\langle \mu^{*(2,2)},\bar q \mu^{(1,0)}+q \mu^{(2,1)}\rangle+\langle \mu^{(2,1)},\bar q \mu^{(1,1)}\rangle + \langle g^{*(3,3)},q\mu^{(1,0)}\rangle \notag\\
     &\, +\dots
 \end{align}
  reproduces the linear equations of motion $\bar\partial g^{(0,0)} + \text{div} \mu^{(1,1)}=0$ and $\bar\partial \mu^{(1,1)} +\text{div}\mu^{*(2,2)}=0$, as well as the $Q$-gauge transformations for $\mu^{(1,1)}$ and $g^{(0,0)}$. The $\cdots$ represent more terms involving antifields corresponding to gauge-for-gauge transformations. This is free BCOV extended by antifields, hence it is a BFV theory \cite{Batalin:1983pz}, with the even Poisson bracket
  \begin{equation}
        \{\mu^{i_1,j_1}(z_1), \mu^{i_2,j_2}(z_2)\} = \delta^{i_1+i_2,3}\delta^{j_1+j_2,3} \delta(z_1-z_2)\Upsilon(\bar\theta^3)\Upsilon(\bar\psi^3)\,.
    \end{equation}
  However, not surprisingly, with this kinetic term the natural potential 
  \begin{align}
     S_{int}= \frac{1}{6}\left(\mathcal{A},\frac{1}{\gamma}\mathcal{A}\wedge \mathcal{A}\right)\sim \frac{1}{6}\langle \mu^{(1,1)},\mu^{(1,1)}\wedge\mu^{(1,1)}\rangle+\cdots
  \end{align}
  does not reproduce the Kodaira-Spencer interaction since $\mu^{(2,1)}$ does not appear in the potential. Indeed, upon  variation w.r.t. $\mu^{(1,1)}$ {and hitting the resulting equation with $\text{div}$} one finds  
  \begin{align}
     \text{div} \bar q \mu^{2,1}+ \frac{1}{2}\text{div} (\mu^{(1,1)}\wedge\mu^{(1,1)})=0\,,
  \end{align}
  but there is no equation relating  $\mu^{2,1}$ to $\mu^{1,1}$.

\subsection{Short multiplets and Kodaira--Spencer with the compensator}
\label{short-multpl}
To get around the obstruction with the interaction term we consider instead two $\bar q$-multiplets $\tilde A$ and $A$, rather than a $Q$-multiplet, alongside with their gauge and gauge for gauge fields and other multiplets that will later play the role of antifields,
  \begin{align}
 &\left(\frac{1}{\gamma^2} g^{(3,0)}\right)\stackrel{
 \bar q}{\to}\left(\frac{1}{\gamma^2} g^{(3,1)}+\frac{1}{\gamma}{\mu^{(2,0)}}\right)\stackrel{
 \bar q}{\to} \left(\frac{1}{\gamma^2} g^{(3,2)}+\frac{1}{\gamma}{\mu^{(2,1)}}+{\mu^{(1,0)}}\right)\stackrel{X}{\to}\\&\left(\frac{1}{\gamma^2}g^{*(3,3)}+\frac{1}{\gamma}{\mu^{*(2,2)}}+{\mu^{(1,1)}}+\gamma g^{(0,0)}\right)\stackrel{\bar q}{\to}\left(\frac{1}{\gamma}{\mu^{*(2,3)}}+{\mu^{*(1,2)}}+\gamma g^{*(0,1)}\right)\stackrel{\bar q}{\to}\left({\mu^{*(1,3)}}+\gamma g^{*(0,2)}\right)\notag\\
 &\stackrel{\bar q}{\to}\left(\gamma^2 g^{*(0,3)}\right), \nonumber  \end{align}
with short hand notation
 \begin{align}
 \tilde A^{gg}\stackrel{\bar q}{\to}\tilde A^g\stackrel{\bar q}{\to}\tilde A\stackrel{X}{\to}A\stackrel{\bar q}{\to}\tilde A^*\stackrel{\bar q}{\to}A^{*g}\stackrel{\bar q}{\to}\tilde A^{*gg}\,,
 \end{align}
where $\tilde A^g$ stands for the multiplet of gauge symmetries for $\tilde A$. 
This truncation will be consistent after using the equation of motion for $\tilde A$, as in section \ref{shs-BVint}, implying $A$ to be in the kernel of $X$.  Then, it is consistent to set  $g^{(3,3)}$ to zero in $A$ since it cannot be in the image of $X$. However, that implies that the pairing is degenerate, since  $g^{(3,3)}$ is the antifield of  $g^{(0,0)}$. We will return to this when analyzing the equations of motion below.

Treating $\tilde A$ as an auxiliary multiplet (or multiplet of potentials) as in section \ref{shs-BVint}, we then choose the action 
\begin{align}\label{eq:Action}
  S(\tilde A,A)&=  \frac{1}{2}\left(\tilde A ,\bar q\,X \tilde A \right)+\left( \tilde A ,\bar q  A \right)+\frac{1}{6}\left(A,\frac{1}{\gamma}A\wedge A\right)\nonumber\\
  & = \frac{1}{2} \langle {\mu}^{(2,1)}, \bar q\, q {\mu}^{(2,1)}\rangle + {\langle {g}^{(3,2)} , \bar q \, q  \mu^{(1,0)}\rangle} +
\langle {\mu}^{(2,1)}, \bar q \mu^{(1,1)}\rangle + \langle {g}^{(3,2)} , \bar q g^{(0,0)}\rangle + \langle \mu^{(2,2)}, \bar q \mu^{(1,0)}\rangle \notag \\
& \; \; \;  + \frac{1}{6}\langle \mu^{(1,1)}, \mu^{(1,1)}\wedge \mu^{(1,1)}\rangle + \langle \mu^{(2,2)}, \mu^{(1,1)}\wedge g^{(0,0)}\rangle{+\frac 1 2 \langle g^{(3,3)}, g^{(0,0)}\wedge g^{(0,0)}\rangle}\, .
\end{align}
Since $\bar q$ and $X$ are both odd operators on the large Hilbert space, the first term above is necessarily even. In order for the action to have homogeneous parity we then need to assign even parity to $A$, which, in turn implies that $\tilde A$ has odd parity etc.   

The equation of motion {for $\tilde{A}$} following from \eqref{eq:Action} breaks down into
\begin{align}
    \bar q \text{div} \mu^{(2,1)} +\bar q \mu^{(1,1)}&=0\,, \label{1st}\\
     \bar q \text{div} \mu^{(1,0)} +\bar q g^{(0,0)}&=0\,, \label{2nd}\\
   \bar q \text{div}{g}^{(3,2)} +\bar q \mu^{(2,2)}&=0\,. \label{3rd}
   \end{align}
We can now repeat the discussion below \eqref{eq:KSsmall}: If $A$ has no projection on the cohomology of $q$, the  equation for $A$ can be solved to express $A=-X\tilde A$, implying, in particular, that $A$ is divergence-free. If, on the other hand, $A\in H_{\bar\partial}(PV^{})$ then it follows (existence of a quasi-iso to $H_{\bar\partial}(\ker \text{div})$ guaranteed by the $\partial\bar\partial$-lemma) that $A$ can be assumed to be divergence free.

Varying w.r.t.~$A$ yields instead 
   \begin{align}
    \bar q \mu^{(2,1)} +\frac{1}{2}\mu^{(1,1)}\wedge \mu^{(1,1)} + \mu^{(2,2)}\wedge g^{(0,0)}&=0\,, \label{4th}\\
     \bar q g^{(3,2)}+ \mu^{(2,2)}\wedge \mu^{(1,1)}+g^{(3,3)} \wedge g^{(0,0)}&=0\,,\label{6th}\\
   \bar q \mu^{(1,0)} + \mu^{(1,1)}\wedge g^{(0,0)}&=0\,,\label{5th}\\
   g^{(0,0)}\wedge g^{(0,0)} &=0  .
\end{align}
{The last equation follows from variation w.r.t. to $g^{(3,3)}$ which does not have a kinetic term) }

 Upon left action by $\text{div}$ on \cref{4th,6th,5th} and using \cref{1st,2nd,3rd}, we find 
\begin{align}\label{eq:cOM}
     \bar q \mu^{(1,1)} +\frac{1}{2}\text{div} (\mu^{(1,1)}\wedge \mu^{(1,1)}) + \text{div}(\mu^{(2,2)} \wedge g^{(0,0)})=0 , \notag \\ 
     \bar q \mu^{(2,2)} + \text{div} (\mu^{(2,2)} \wedge \mu^{(1,1)})=0,\notag \\
     \bar q g^{(0,0)} + \text{div}(\mu^{(1,1)} \wedge g^{(0,0)})=0\,,
\end{align}
which agrees with \eqref{eq:ELCostello}, \eqref{eq:ELCostello2} obtained by Costello and Li \cite{Costello:2019jsy} for divergence free polyvectors. Let us now return to $g^{(3,3)}$ which is a singlet so that we have the freedom to include it or not. If we include it, {\cref{1st,2nd,3rd} is a symplectic reduction and the $g^{(3,3)}$} variation implies the equation of motion 
\begin{align}
   {g^{(0,0)}\wedge g^{(0,0)}}&{=0\,,}\label{7th}
\end{align}
and thus $g^{(0,0)}=0$, so that we recover the result in the small Hilbert space. If we exclude it, the symplectic form on the reduced space is degenerate and the reduction is coisotropic. We follow the latter option here.

Now we would like to make contact with section \ref{shs-BVint} and \cite{Barannikov:1997dwc,Costello:2012cy,Costello:2019jsy}. Hence we use \cref{1st,2nd,3rd}, that follow from variation w.r.t. to the auxiliary fields in $\tilde A$, and substitute $A$ in $S(A,\tilde A)$ as $A=-X\tilde A+h$, with $h$ a compactly-supported harmonic term, to get 
\begin{align}
    -S(\tilde A,h)=&\frac{1}{2}(\tilde A,\bar q X\tilde A)+\frac{1}{6}\left(X\tilde A-h,\frac{1}{\gamma}(X\tilde A-h)\wedge (X\tilde A-h)\right)\nonumber\\
    =&\frac{1}{2}\langle  \mu^{(2,1)},\bar q q  \mu^{(2,1)}\rangle+\langle  g^{(3,2)},\bar q q \mu^{(1,0)}\rangle+\langle q g^{(3,2)}{+h^{(2,2)}},(q \mu^{(2,1)}+{h^{(1,1)}})\wedge q \mu^{(1,0)} + h^{(0,0)}\rangle \notag \\
    & +\frac{1}{6}\langle q  \mu^{(2,1)}+{h^{(1,1)}}, (q\mu^{(2,1)}+{h^{(1,1)}})\wedge (q \mu^{(2,1)}+{h^{(1,1)}})\rangle  .
    \label{BK-with_g}
\end{align}
This is then the extension of the action in \cite{Barannikov:1997dwc} to include the compensator field $g$. Invariance of $S$ under the gauge symmetries for $\tilde A$ dictates that the minimal BV extension, including the gauge symmetries for the gauge parameter $\tilde{A}^g$ and antifields\footnote{For convenience, we leave the respective harmonic terms out of our considerations.}, is given by 
  \begin{align}
    &S(\tilde A,h)- (\tilde A^*, \bar q\tilde A^g+\frac{1}{
  \gamma} X \tilde A^g\wedge (X\tilde A-h)
  ) \notag \\
  &\, -(\tilde{A}^{*g}, \bar q \tilde A^{gg}+{\frac{1}{2\gamma}} X\tilde{A}^g \wedge X\tilde{A}^g {{+ \frac 1 \gamma }X\tilde{A}^{gg} \wedge (X \tilde A-h) }) \notag \\
  & \, - (\tilde{A}^{*gg}, \frac 1 \gamma X\tilde{A}^g \wedge X\tilde{A}^{gg})=: \mathcal{S}\, .
  \label{action_wghosts}
\end{align}
Parity of $\vert \mathcal{S}\vert$ fixes the parity of the antifields, so $\tilde A^* $ and $\tilde A^{* gg}$ are odd, while $\tilde A^{*g}$ is even. The gauge transformations $\delta\tilde A = \frac{\delta S}{\delta \tilde A^*}$ can be unpacked as 
\begin{align}
    -\delta  \mu^{(2,1)}&=\bar q  \mu^{(2,0)}+q  \mu^{(2,0)} \wedge (q \mu^{(2,1)} - h^{(1,1)}) {{+  q g^{(3,1)}\wedge (q \mu^{(1,0)}-h^{(0,0)})}}, \\
    -\delta  \mu^{(1,0)}&=(q \mu^{(1,0)}-h^{(0,0)}) \wedge q  \mu^{(2,0)}, \\
    -\delta  g^{(3,2)}&={\bar q g^{(3,1)}+} (q  g^{(3,2)} - h^{(2,2)}) \wedge q  \mu^{(2,0)}\,. 
    \label{gauge_of-BK-with_g}
\end{align}
Upon left action by $q$ and using \cref{1st,2nd,3rd}, as well as defining
\[
\lambda^{(1,0)}\equiv -q\mu^{(2,0)},\quad \lambda^{(2,1)} \equiv -q g^{(3,1)}\, ,
\]
these take the more intuitive and familiar form 
\begin{align}
    \delta  \mu^{(1,1)}&= \bar q  \lambda^{(1,0)} +[ \lambda^{(1,0)}, \mu^{(1,1)}]+ [\lambda^{(2,1)}, g^{(0,0)}], \nonumber\\
    \delta g^{(0,0)}&=[g^{(0,0)} , \lambda^{(1,0)}], \\
    \delta \mu^{(2,2)}&= \bar q \lambda^{(2,1)}+[\mu^{(2,2)} ,\lambda^{(1,0)}] \,. \nonumber
    \label{gauge_of_BK-with-g abridged}
\end{align}
Similarly, the third line in \eqref{action_wghosts} encodes the gauge for gauge symmetries: The (even) ghost multiplet $\tilde A^g$ transforms as
\begin{align}
    & { -\delta \mu^{(2,0)} = -\frac 1 2 q \mu^{(2,0)} \wedge q \mu^{(2,0)} + q g^{(3,0)} \wedge (q \mu^{(1,0)}}+ h^{(0,0)}), \\
    &{-\delta g^{(3,1)} =\bar q g^{(3,0)}} + qg^{(3,1)} \wedge q \mu^{(2,0)} + { q g^{(3,0)} \wedge (q \mu^{(2,1)}}+h^{(1,1)}).
\end{align}

We still need to identify the corresponding BV-bracket and show that \eqref{action_wghosts} is a solution to the a classical master equation. Let us inspect the reduced theory. If $a\in \tilde A , \tilde A^g , \tilde A^{gg} \subset PV^{i,j}\gamma^k$ with 
$(i,j)\in \{ (1,0), (2,0), (3,0), (2,1), (3,1), (3,2)\}$, then its conjugated is $a^{*} \in \tilde A^*, \tilde{A}^{*g}, \tilde{A}^{*gg}\subset \text{div} PV^{4-i,3-j} \gamma^{-k-1}$.\footnote{Note that the first degree makes sense, since $i\geq 1$.} So we choose the bracket: 
\begin{equation}
           \{S,F\} = \left.\left(\frac{\delta S}{\delta a} \frac{\delta F}{\delta a^{*}} + (-1)^{\vert a \vert \, \vert a^*\vert} \frac{\delta S}{\delta a^{*}} \frac{\delta F}{\delta a} \right) \right|_{{a^{*} \in \text{div} PV }} \; 
        \delta^{(3)}(z-z') \Upsilon(\bar\theta^3)\Upsilon(\bar\psi^3)
        \label{symplform}
    \end{equation}
Acting with both functional derivatives does not change the ghost degree, but constrained variation in the div exact subspace does. Then it becomes an odd bracket after projection, akin to BCOV's odd Poisson bracket. 

Note that the datum $\tilde A^{gg} \xrightarrow{\bar q}\tilde A^{g} \xrightarrow{\bar q}\tilde A \xrightarrow{\sim}\tilde A^{*} \xrightarrow{\bar q}\tilde A^{*g} \xrightarrow{\bar q}\tilde A^{*gg} $ can now be reinterpreted as a BV complex for each component field of $\tilde A$
            \begin{align*}
  \begin{matrix}
    gh: &3 & &2 & & 1 & & -1 & & -2 & & -3 \\
   &0 &\rightarrow  & 0 &\rightarrow & \mu^{(1,0)} &\xrightarrow{\sim} & \frac{ 1 }{\gamma}\mu^{*(2,3)} &\rightarrow & 0 &\rightarrow & 0\, ,\\
   & 0 &\rightarrow  & \frac{1}{\gamma} \mu^{(2,0)} &\xrightarrow{\bar q} &  \frac 1 \gamma \mu^{(2,1)} &\xrightarrow{\sim} & \mu^{*(1,2)} &\xrightarrow{\bar q} &  \mu^{*(1,3)}&\rightarrow & 0\,,\\
    & {\frac{1}{\gamma^2} g^{(3,0)}} &\xrightarrow{\bar q}  & {\frac{1}{\gamma^2} g^{(3,1)}} &\xrightarrow{\bar q} & \frac{1}{\gamma^2}\mu^{(3,2)} &\xrightarrow{\sim} & \gamma \mu^{*(0,1)}&\xrightarrow{\bar q} & \gamma \mu^{*(0,2)}&\xrightarrow{\bar q} & \gamma \mu^{*(0,3)} \, .
     \end{matrix}
    \end{align*}
Eventually the BV extension $\mathcal{S}$ satisfies the CME, \[ \{\mathcal{S},\mathcal{S}\} = 0.\] 
The calculation is straightforward and the conclusion follows from $q$-exactness of the antifield, $\langle q \,a, b\rangle = -(-1)^{\vert a \vert} \langle a, q\,b\rangle$ and the usual compatibility conditions of the differential $\bar q$ with the Schouten bracket (the latter expressed in Bogomolov--Tian--Todorov's version), as well as its Jacobi identity. 

We close this section with a comment on higher order vertices. Interactions beyond cubic order could be included via higher powers of $g^{(0,0)}$ which correspond to higher deformations of the holomorphic form (via $e^{(g^{(0,0)})}$).

\section{Theory of background fields}
\label{sec:backgr-fields}
In the previous section we derived a local and polynomial BV-action for the Kodaira--Spencer theory of complex structure deformations on a Calabi--Yau complex 3-fold. However, since Kodaira--Spencer theory is well posed for any complex manifold, one may wonder why the additional CY-structure is needed for the worldline. In this section we present an alternative, background independent formulation, where we absorb the Beltrami differential into a deformation of the differential $Q$.  
The obvious deformation is the ``minimal coupling" 
\begin{align}
    \bar p_a\to \bar p_a+\mu_{ a}^bp_b\equiv (\bar p_\mu)_a
\end{align}
with $p_a,\bar p_a\sim \partial_{z^a}, \partial_{\bar z^a}$ on the representation space  $V$. Now, what is the correct condition we should impose on the deformed $\bar q$ for the background to be Maurer--Cartan? We will say that a deformation of an almost complex structure is MC if the problem of infinitesimal deformations around that background  is well-defined, i.e.~it can be formulated as a cohomology problem to distinguish between fake (reparametrizations) and actual infinitesimal deformations. For this, $\bar q$ needs to be nilpotent on the relevant vector space spanned by infinitesimal deformations, which is tantamount to 
\begin{align}
    [(\bar p_\mu)_a,(\bar p_\mu)_b]=0,
\end{align}
which is just the KS-equation for integrable deformations of complex structures on a complex manifold. See also \cite{Bershadsky:1994sr} for an earlier discussion. Note that {since a pairing on $V$ is not required here, then consequently there is no need for the compensator multiplet $g$ for the latter. Thus, }the Calabi--Yau condition is not required here. However, a locally invertible holomorphic $3$-form is required to establish an ``operator-state-correspondence" between states in $V$ and infinitesimal deformations of $\bar q$ (see e.g.  \cite{Bockisch:2022eas}). Indeed, in order to produce a state by action of $\bar q_\mu = \bar\psi \cdot\bar p_\mu$, we need a vector space with highest weight vector  the holomorphic volume form $\Upsilon_{abc} \theta^a\theta^b \mathrm{d}z^c$. Then:
\[
\bar q_\mu \Upsilon_{abc} \theta^a\theta^b \mathrm{d}z^c = \bar\psi^d \, \mu_d{}^c(z) \, \Upsilon_{abc}\theta^a\theta^b , 
\]
so the differential operator corresponds to the Beltrami.

\section{Outlook and conclusions}
In this paper we gave a construction of a BV action for BCOV starting from a spinning worldline with local supersymmetry instead of a topological sigma-model, mimicking familiar constructions from string field theory. Working in the small Hilbert space this gives a direct construction of the BV-extension of the action of Barannikov and Kontsevich \cite{Barannikov:1997dwc} as a theory of potentials (plus harmonic polyvector fields. 

In the large Hilbert space formulation we constructed a multiplet that includes the scalar compensator for the holomorphic 3-form among other gravitational descendents which naturally arise as the result of a $U(1)$ action on the superghost $\gamma$. Compared to Costello--Li's formulation \cite{Li:2014nem}, the off-shell action constructed here is formulated in terms of potentials (plus harmonic polyvectors) and this  includes a local kinetic term. 

Concerning the question of the gauge symmetries, in the non-linear theory with the local kinetic term (either in the large or small Hilbert space), after using the field equations to project the field onto its div-exact and harmonic part, we retrieve the familiar symmetry for the Beltrami differential (adjoint action of the vector field for holomorphic diffeos).

In the small Hilbert space formulation, with an arguably more geometric interpretation in terms of pictures, we are able to formulate Kodaira--Spencer theory just for complex structure deformations without inclusion of the compensator for the holomorphic $3$-form. For that we needed to resort to the large Hilbert space with the drawbacks just described.  However, it is not obvious that the compensator could be included in the small Hilbert space as well if one were to consider reducible representations (BV-multiplets). A posteriori, this may not be so surprising since only in the large Hilbert space there is an extra $U(1)$-symmetry (phase of $\gamma$) that gives rises to equivariant cohomology while in the small Hilbert space  formulation $\gamma$ is real.

While geometrizing the non-standard kinetic term of BCOV, we believe that our description also highlights Kodaira--Spencer gravity as a valuable toy model to explore and geometrize peculiar features of super string field theory such as picture changing and transition between the small and large Hilbert spaces. In particular, we gave a new interpretation of the pairing in the large Hilbert space. Also, it would be interesting to explore the relation between picture changing and the equivariant differential in \cite{Li:2014nem}. 

The worldline description has the advantage that the superghost sector, which plays a central role in our investigation, is readily included in the operator formulation while in a worldsheet sigma-model its implementation is less direct, in terms of bosonized ghost, which obscures its geometric interpretation. It may thus be instructive to revisit our construction for the world sheet sigma model with a topological twist.

{Our considerations were purely classical, so it is natural to wonder about the quantization of this BV theory. There is already an extensive literature on quantum BCOV theory, including  \cite{Bershadsky:1993cx, Costello:2012cy,Li_2023}. Our approach may open the way to construct the quantum theory directly from the world graphs (including loops) of the underlying $(2;2)$-particle model. For this a description in terms of  tropical geometry may be appropriate  \cite{Tourkine:2013rda,losev2023tropicalmirrorsymmetrycorrelation}.}

\section*{Acknowledgments} 
E.B.~and I.S.~would like to express deep gratitude for the program ``Cohomological aspects of Quantum Field Theory" 2025, at Mittag-Leffler Institute in Stockholm, supported by the Swedish Research Council under grant no. 2021-06594. Part of this work was developed there, profiting from a vibrant, inspiring environment. E.B.~is especially grateful to D. Fiorenza and S. Ronchi for discussions on differential geometry and aspects of the Batalin-Vilkovisky formalism. We would also like to thank Carlo Cremonini and Maxim Zabzine for helpful discussions, and the SciPost referees who suggested several significant improvements. I.S.~ would like to thank the  Theory group at CERN for hospitality during part of this work.  
E.B.~acknowledges support from GA\v{C}R grant PIF-OUT 24-10634O. I.S.~was supported in parts by the Excellence Cluster Origins of the DFG under Germany’s
Excellence Strategy EXC-2094 390783311 as well as EXC 2094/2: ORIGINS 2. Furthermore, this research was supported in part by grant NSF PHY-2309135 to the Kavli Institute for Theoretical Physics (KITP).
\appendix

\section{Deformation problem of complex structures}
\label{App}

We present a short reminder about the deformation problem of complex structures following \cite{Schnell}. Given a complex structure $J$ on a manifold $M$, a splitting between holomorphic and antiholomorphic vector fields is in place. If we are provided with a new (almost) complex structure $J'$, sufficiently close to $J$, then the projection of $TM'$ to its antiholomorphic component will induce an isomorphism and thus we end up with the chain of isomorphisms:
\[
TM_{(1,0)} \xrightarrow{\pi_{(0,1)}^{-1}} TM'_{(1,0)} \xrightarrow{\pi_{(1,0)}} TM_{(0,1)} \, .
\]
In the second slot, we must think of $TM'_{(1,0)}$ as $TM'_{(1,0)} \subset TM'\otimes \mathbb{C}$.
Therefore, there exists a global antiholomorphic 1-form, with values in holomorphic tangent vectors, the \emph{Beltrami differential}:
\begin{align}
\mu \in \Omega^{(0,1)}(M, TM_{(1,0)}) \, .
\end{align}
We then seek the conditions on $\mu$ so to promote the new almost complex structure to a complex structure. Newlander--Nirenberg theorem states that an almost complex structure is fully-fledged complex if it is integrable:
\[
[X,Y] +J'( [J'X,Y] + [X,J'Y]) - [J'X,J'Y] = 0, \quad \forall \, X,Y \in \mathfrak{X}(M).
\]
Equivalently, $TM'_{(0,1)}$ is an involutive distribution.
This reflects in the following Maurer--Cartan equation for $\mu$:
\begin{equation}
    \bar\partial \mu + \frac{1}{2} [\mu, \mu] = 0 \, ,
    \label{MaurerCartan}
\end{equation}
where $[-,-]$ is the Lie bracket of vector fields (the reader should be reminded that it extends on multivector fields as the Schouten--Nijenhuis bracket). There is another equivalent expression to \eqref{MaurerCartan} based on Bogomolov--Tian--Todorov lemma \cite{Tian1987SmoothnessOT} (see also \cite{rao2019tiantodorovlemmaapplicationsdeformation} for a review), which we will display later. First let us recall that the holomorphic Dolbeault differential on forms induces a differential operator $\text{div}$ on polyvector fields. The latter are forms that take values in multivector fields, i.e. \[
PV^{\bullet,\bullet}(M) := \Omega^{0,\bullet}(M , \Lambda^\bullet TM_{(1,0)}) \, .
\]
The differential operator $\text{div}$ is then defined by the commutative diagram:
\begin{equation}
    \begin{tikzcd}
PV^{\bullet,\bullet} \arrow{r}{\text{div}} \arrow[swap]{d}{\overset{\Upsilon}{\sim}} & PV^{\bullet-1,\bullet} \arrow{d}{\overset{\Upsilon}{\sim}} \\%
\Omega^{d-\bullet,\bullet} \arrow{r}{\partial} & \Omega^{d+1-\bullet, \bullet}
\end{tikzcd}\,.
\label{comm_diagram}
\end{equation}
Here $\partial$ is the holomorphic Dolbeault differential and $\Upsilon$ is the non-degenerate holomorphic volume form thanks to which $PV^{k,m} \cong \Omega^{d-k,m}$. Such an element exists only for the case of Calabi--Yau manifolds. So \cite{Tian1987SmoothnessOT}:
\begin{equation}
[\alpha,\beta]_{SN} \equiv \text{div} (\alpha \wedge \beta) - (\text{div}) \wedge \beta - (-1)^{\vert \alpha \vert} \alpha \wedge \text{div} \beta,
\label{TIAN}
\end{equation}
where the wedge product is intended to be taken on both the form part and the multivector fields.
Eventually, on div-free Beltrami differentials, \eqref{MaurerCartan} can be rewritten as:
\begin{equation}
\bar\partial \mu + \frac 1 2 \text{div} (\mu \wedge \mu) = 0, \quad \mu \in {{\ker \text{div} \subset}} \, PV^{1,1}(M) \, .
\label{Tian-version}
\end{equation}

Complex structure deformations of Calabi--Yau manifolds may or may not preserve the volume form. If $g\in PV^{(0,0)}$ parametrizes the freedom in choosing a global factor for $\Upsilon$, then to preserve the holomorphic volume form one has to ask that $\iota_{\exp g \exp \mu} \Upsilon$ is closed. This yields:
\begin{align*}
 0= \mathrm{d}(\iota_{\exp g \exp \mu} \Upsilon) = &\mathrm{d} (\iota_{\exp g}\iota_{\exp \mu} \Upsilon) = \mathrm{d} \iota_{g} \Upsilon + \mathrm{d} \iota_{\mu} \Upsilon -\iota_\mu \mathrm{d}(\iota_g \Upsilon) + \iota_g \mathrm{d}(\iota_\mu \Upsilon) + \mathrm{d} (\iota_g \iota_\mu \Upsilon) + \mathcal{O}(3) \\
  = & \left(\iota_{\bar\partial g} + \iota_{\text{div} \mu} + \iota_{[g,\mu]}\right)  \Upsilon + \mathcal{O}(3) .
\end{align*}
In conclusion, complex structure deformations that also preserve the holomorphic volume form on a Calabi--Yau manifold ought to satisfy:
\begin{equation}
    \bar\partial g + \text{div}\mu + [g,\mu] = 0.
\label{g-compensates-vol}
\end{equation}

\section{A fully gauged $(2;2)$-worldline}
\label{BRST-22}

Alternatively, we could motivate our choice for the operator algebra \eqref{eq:wlws} for BCOV theory by starting with the spinning particle whose worldline has $\mathcal{N}=(2;2)$-supersymmetry. Then BRST quantization gives rise to the differential 
\begin{align}
    Q=c\mathrm{H}+\gamma q +\gamma^\dagger q^\dagger+\bar\gamma \bar q +\bar\gamma^\dagger \bar q^\dagger + b(\gamma\gamma^\dagger+\bar\gamma\bar\gamma^\dagger) .
    \label{BRSTdifferential}
\end{align}
Written in Darboux coordinates, the $2 + 2$ supercharges in \eqref{BRSTdifferential} take the form
\[
\bar q^\dagger =\psi \cdot p , \quad \bar{q} = \bar\psi \cdot \bar p , \quad q = \theta \cdot p , \quad {q}^\dagger = \bar\theta \cdot \bar p \, .
\]
with 
\begin{equation}
\{\psi^a,\bar\psi^{\bar b}\}= h^{a\bar b} = \{\theta^a,\bar\theta^{\bar b}\}\, 
\implies \{\bar q, \bar{q}^\dagger\}= \mathrm{H} = \{q, {q}^\dagger\},
\label{algebra}
\end{equation}
provided the target space is K\"ahler.\footnote{As matter of fact, it would be appropriate to deploy covariant momenta and thus Christoffel symbols in the supercharges, but we will not need their explicit expressions in the following therefore we spare the effort.} 
All other brackets are zero. $\mathrm{H}$ is the worldline Hamiltonian whose explicit form will not be relevant here, as it relates to diffeomorphism invariance, a non-topological feature that we are going to lift. Indeed, 
in order to make contact with the topological string, we shall project out all information about the punctures on super moduli space $\mathcal{M}$ of the $\mathcal{N}=(2;2)$ worldline. For this, we
choose the path integral measure to produce a constant function (rather than a top form) on $\mathcal{M}$. Essentially, one does not want to know where the punctures are.

This is reasonable, because there is a canonical way to write interaction terms on the worldsheet (or worldline) whereas the kinetic terms involve some choices. Our choices will be as follows. We first eliminate the Hamiltonian constraint in $Q$ by going from  \begin{align}\label{eq:iso2}
 V=C^\infty(U)[\bar\theta^1, \dots \bar\theta^n\vert\bar\psi^1, \dots \bar\psi^n] \otimes \mathbb{C}[[c,\gamma,\bar\gamma,\gamma^\dagger,\bar\gamma^\dagger]] \cong \Lambda^\bullet (TU_{(1,0)}[1]\oplus T^*U_{(0,1)}) \otimes \mathbb{C}[[c,\gamma,\bar\gamma,\gamma^\dagger,\bar\gamma^\dagger]]
\end{align}
where $U \subset CY$, to the reduced module
\begin{align}
  {{V_{red}}} :=  {{{H}}}_{b\gamma\gamma^\dagger}\cap {{H}}_{b\bar\gamma\bar\gamma^\dagger}=V/\{\gamma\gamma^\dagger,\bar\gamma\bar\gamma^\dagger\}\,,
  \end{align} 
where {the Hamiltonian} $\mathrm{H}$ is exact. 
This reduction does not yet delete all dependence on supermoduli but this can be achieved by a presymplectic formulation with a suitable degenerate symplectic form. 

After this kick-starter, we shall now suggest a pairing. 
One possible presymplectic formulation amounts to simply insert $\delta$-functions for all the ghosts which would be the canonical procedure for the interaction term. However, this completely trivalizes $Q$ inside expectation values. Therefore, {in the notation of the body of this article,}  we define the pairing by 
\begin{align}\label{eq:pair}
\left(a,b\right)=\int_{CY^3} \mathrm{d}vol_{CY^3} 
\int \mathrm{d}\gamma\mathrm{d}\bar\gamma\mathrm{d}\gamma^\dagger\mathrm{d}\bar\gamma^\dagger < a\, b \,  
\Upsilon(\theta^{\wedge 3})\Upsilon(\psi^{\wedge 3})>\;\delta(\bar\gamma^\dagger)\delta(\gamma^\dagger)\delta'(\bar\gamma), \, \quad a,b \in V_{red}
\end{align}
in the path integral measure. 
The lack of $\delta(\gamma)$ implies that our pairing has ``picture 1". Note how this pairing has total ghost number given by $gh(\delta(\gamma))$, i.e.~the choice of the ghost number of the vacuum. Moreover, $q=\theta \cdot p$ and $\bar q = \bar\psi \cdot\bar p$ are self-adjoint w.r.t.~the pairing (one way to see this is to use the existing isomorphisms with the module of $PV^{\bullet,\bullet}$ and integration thereof, and note that $\bar q \sim \bar\partial$ as well as $q \sim \text{div}$, which are self-adjoint w.r.t.~the integral.).

That done we can easily describe the structure states in $V_{red}$, modulo terms that will not contribute due to the degeneracy in the symplectic structure. We just need to choose a representation for the remaining ghost algebra $[\beta,\gamma]=1$. There are two inequivalent representations, given by polynomials in $\gamma$ and derivatives of $\delta(\gamma)$.  In the polynomial representation of $V_{red}$ we have 
\begin{equation}
\Phi^{(0)}  = \oplus_{\small p,q}\left(\phi_{0\;b_1\cdots b_q}^{a_1\cdots a_p}(z) \bar\theta_{a_1} \cdots \bar\theta_{a_p}\bar\psi^{ b_1}\cdots \bar\psi^{ b_q} +\bar\gamma  \phi_{\bar 1^\dagger\; b_1\cdots b_q}^{a_1\cdots a_p}(z) \bar\theta_{a_1} \cdots \bar\theta_{a_p}\bar\psi^{ b_1}\cdots \bar\psi^{ b_q}\right) + O(\gamma).
\label{multiplet-of-Beltrami}
\end{equation}
Now, again due to the missing delta function $\delta(\gamma)$ in the pairing  \eqref{eq:pair}, this representation pairs with the ``picture $-1$",
\begin{equation}
\Phi^{(-1)}  =  \oplus_{p,q}\left(\varphi_{\bar 1^\dagger\;b_1\cdots b_q}^{a_1\cdots a_p}(z) \bar\theta_{a_1} \cdots \bar\theta_{a_p}\bar\psi^{b_1}\cdots \bar\psi^{ b_q} \,\bar\gamma\delta(\gamma)+ \varphi_{0\;b_1\cdots b_q}^{a_1\cdots a_p}(z) \bar\theta_{a_1} \cdots \bar\theta_{a_p}\bar\psi^{ b_1}\cdots \bar\psi^{ b_q} \,\delta(\gamma)\right)+ O(\partial_\gamma)\delta(\gamma).
\label{multiplet-of-Beltrami-1}
\end{equation}
This module and its dual are quite large and eventually we can focus only on components given by $a^{(0)}$ \eqref{eq:a0} and $a^{(-1)}$ \eqref{eq:a0*} in the body of this article, which however do not depend on $\bar\gamma$. The correspondence to that half-gauged sigma model is clear if one notes that there is an equivalent formulation to \eqref{eq:pair} where we absorb $\bar\gamma$ 
in the component fields, thus stripping away the $\bar\gamma$-dependence of the states \eqref{eq:a0} and \eqref{eq:a0*}, and work with the pairing 
\begin{align}\label{eq:pairn}
\left(a,b\right)=\int_{CY^3} \mathrm{d}vol_{CY^3}
\int \mathrm{d}\gamma\mathrm{d}\bar\gamma\mathrm{d}\gamma^\dagger\mathrm{d}\bar\gamma^\dagger < a \,b \, 
\Upsilon(\theta^{\wedge 3}) \Upsilon(\psi^{\wedge 3})>\;\delta(\bar\gamma^\dagger)\delta(\gamma^\dagger)\delta(\bar\gamma))
\end{align}
with an un-differentiated $\delta(\bar\gamma)$ and $\bar\gamma \bar q \to \bar q \sim \bar\partial$.

\bibliographystyle{ieeetr}
\bibliography{sample}

\end{document}